\documentclass[aps,prb,twocolumn,superscriptaddress]{revtex4-2}
\usepackage{graphicx}
\usepackage{mathrsfs}
\usepackage{xcolor}
\usepackage[colorlinks=true,allcolors=blue,citecolor=blue]{hyperref}
\usepackage{bm}
\usepackage{amsmath}
\usepackage{dcolumn}
\usepackage{epstopdf}
\usepackage{dsfont}
\usepackage{amssymb}
\usepackage{tabularx}
\usepackage{array}
\usepackage{colordvi}
\usepackage{braket}
\usepackage{float}
\usepackage{mathtools}
\usepackage{upgreek}
\usepackage{booktabs}
\usepackage{multirow}
\usepackage{threeparttable}
\usepackage{appendix}
\usepackage{fixltx2e}

\begin{document}
\title{Magnetic order-dependent giant tunneling magnetoresistance and electroresistance in van der Waals antiferromagnetic-multiferroic tunnel junctions}
\author{Zhi Yan$^\S$}
\email[Corresponding author:~]{yanzhi@sxnu.edu.cn}
\affiliation{Research Institute of Materials Science $\&$ Key Laboratory of Magnetic Molecules and Magnetic Information Materials of Ministry of Education $\&$ Colege of Physics and Information Engineering, Shanxi Normal University, Taiyuan 030031, China}
\affiliation{School of Chemistry and Materials Science $\&$ Collaborative Innovation Center for Shanxi Advanced Permanent Magnetic Materials and Technology, Shanxi Normal University, Taiyuan 030031, China}
\thanks{These authors contributed equally to this work.}
\author{Dan Qiao$^\S$}
\affiliation{Research Institute of Materials Science $\&$ Key Laboratory of Magnetic Molecules and Magnetic Information Materials of Ministry of Education $\&$ Colege of Physics and Information Engineering, Shanxi Normal University, Taiyuan 030031, China}
\thanks{These authors contributed equally to this work.}
\author{Wentian Lu}
\affiliation{Research Institute of Materials Science $\&$ Key Laboratory of Magnetic Molecules and Magnetic Information Materials of Ministry of Education $\&$ Colege of Physics and Information Engineering, Shanxi Normal University, Taiyuan 030031, China}
\author{Xinlong Dong}
\email[Corresponding author:~]{dongxl@sxnu.edu.cn}
\affiliation{Research Institute of Materials Science $\&$ Key Laboratory of Magnetic Molecules and Magnetic Information Materials of Ministry of Education $\&$ Colege of Physics and Information Engineering, Shanxi Normal University, Taiyuan 030031, China}
\author{Xiaohong Xu}
\email[Corresponding author:~]{xuxh@sxnu.edu.cn}
\affiliation{Research Institute of Materials Science $\&$ Key Laboratory of Magnetic Molecules and Magnetic Information Materials of Ministry of Education $\&$ Colege of Physics and Information Engineering, Shanxi Normal University, Taiyuan 030031, China}
\affiliation{School of Chemistry and Materials Science $\&$ Collaborative Innovation Center for Shanxi Advanced Permanent Magnetic Materials and Technology, Shanxi Normal University, Taiyuan 030031, China}

\date{\today{}}

\begin{abstract}
Antiferromagnetic spintronics exhibits ultra-high operational speed and stability in a magnetic field, holding promise for the realization of next-generation ultra-high-speed magnetic storage. However, theoretical exploration of the electronic transport properties of antiferromagnetic-multiferroic tunnel junction (AMFTJ) devices remains largely unexplored.
Here, we design an antiferromagnet/ferroelectric barrier/antiferromagnet van der Waals heterojunction, renamed vdW AMFTJ, using a bilayer MnBi$_2$Te$_4$/In$_2$Se$_3$/bilayer MnBi$_2$Te$_4$ (MBT-2L/IS/MBT-2L) as the prototype.
Based on first-principles calculations using the nonequilibrium Green's function method combined with density functional theory, we theoretically investigate the spin-resolved electronic transport properties of this AMFTJ.
By manipulating the various possible magnetization directions of the multilayer antiferromagnetic MnBi$_2$Te$_4$ and the ferroelectric polarization direction of the In$_2$Se$_3$ within the junction, sixteen distinct non-volatile resistance states can be revealed and manipulated by applying external biaxial strain and bias voltage.
We predict maximum tunneling magnetoresistance (electroresistance) values of $3.79\times10^{4}$\% ($2.41\times10^{5}$\%) in the equilibrium state, which can increase up to $5.01\times10^{5}$\% ($4.97\times10^{5}$\%) under external bias voltage. 
Furthermore, the perfect spin filtering effect is also present in our AMFTJ.
Our results highlight the tremendous potential of the MBT-2L/IS/MBT-2L vdW AMFTJ in non-volatile memory, expanding the application avenues for antiferromagnetic spintronic devices.

\end{abstract}

\maketitle
\section{Introduction}
The burgeoning field of spintronics, which exploits the intrinsic spin of electrons along with their charge, has driven the quest for advanced materials and device architectures that can meet the demands for faster, more efficient, and higher density information storage and processing~\cite{wolf2001spintronics,vzutic2004spintronics}. Traditionally, ferromagnetic materials have underpinned spintronic devices such as magnetic tunnel junctions (MTJs)~\cite{zhu2006magnetic,ikeda2007magnetic} and spin-transfer torque (STT) memories~\cite{ralph2008spin,kawahara2012spin}. However, these materials are inherently limited by their susceptibility to external magnetic fields, leading to potential interference and instability in high-density device arrays. This has catalyzed the search for alternative magnetic materials that can overcome these limitations.

Antiferromagnetic materials have recently emerged as strong contenders due to their unique properties~\cite{han2023coherent,dal2024antiferromagnetic}. Characterized by antiparallel spin alignment that results in a zero net magnetic moment, antiferromagnets exhibit immunity to external magnetic fields and ultrafast spin dynamics that operate at terahertz frequencies~\cite{qin2023room,gu2023multi}. These attributes render them particularly suitable for high-speed, low-power spintronic applications. Recent advances have demonstrated the potential of antiferromagnetic spin valves~\cite{zhai2021electrically}, antiferromagnetic STT memory~\cite{chirac2020ultrafast}, antiferromagnetic spin-orbit torque (SOT) devices~\cite{wang2023field}, and antiferromagnetic tunnel junctions~\cite{qin2023room,chen2023octupole}, highlighting the versatility and promise of these materials.

Multiferroic heterostructure systems exhibit both ferroelectric and (anti)ferromagnetic order, introducing an additional controllable dimension through their coupled electric and magnetic properties~\cite{gu2023multi,zhang2024electronic}. This coupling enables the manipulation of magnetic states via electric fields and vice versa, enabling the development of multifunctional devices with enhanced capabilities. Antiferromagnetic-multiferroic tunnel junctions (AMFTJs), which combine antiferromagnetic and ferroelectric layers, are promising candidates for next-generation spintronics devices. They promise high tunneling magnetoresistance/electroresistance (TMR/TER), electric-field control of spin transport, and robustness against magnetic field interference.
Two-dimensional van der Waals (vdW) materials, with atomically thin structures, offer unparalleled opportunities for creating heterostructures with tailored functionalities through precise control of electronic and magnetic properties via layer stacking and external stimuli. This makes them ideal candidate materials for novel spintronic devices~\cite{0dong2023spin,0su2020van,0feng2024van,0yan2020significant,0yang2024multistate,0yu2023fully,0yan2021barrier,0zhang2021recent,0zhu2021giant,0yan2024giant,0zhang2023current,50yan2022giant}. Notably, integrating two-dimensional vdW antiferromagnetic and ferroelectric materials into tunnel junctions remains an uncharted territory, presenting a unique opportunity to explore vdW AMFTJs in this new dimensionality.

In this paper, we introduce a novel vdW AMFTJ device, employing a bilayer MnBi$_2$Te$_4$/In$_2$Se$_3$/bilayer MnBi$_2$Te$_4$ (MBT-2L/IS/MBT-2L) as the prototype and graphite as an electrode. Then we investigate the spin-dependent electronic transport properties associated with 16 distinct resistance states of this vdW AMFTJ through $ab$ $initio$ quantum transport simulations.
Our calculations indicate that TMR and TER exhibit significant non-volatile multiple states driven by the switchable ferroelectric polarization direction in different magnetic configurations. Remarkably, we also observe perfect spin filtering effects under special magnetic configurations and finite bias voltages.
In addition, the electronic transport properties of the AMFTJ can be flexibly tuned by an external bias voltage and biaxial strain. Under the influence of bias voltage, the maximum TMR (TER) achieved is $5.01\times10^{5}$\% ($4.97\times10^{5}$\%).
Our findings provide critical insights into the design and optimization of two-dimensional vdW AMFTJs, laying the foundations for their integration into advanced spintronic devices with enhanced functionalities and reduced power consumption.

\section{COMPUTATIONAL Methods}
The structural relaxation, total energy, and electronic band structure calculations were performed using density functional theory within the Vienna $Ab$ $initio$ Simulation Package (VASP)~\cite{31kresse1996efficient}. 
We employed the projector-augmented wave pseudopotentials method~\cite{32blochl1994projector} and the generalized gradient approximation (GGA) in the Perdew-Burke-Ernzerhof (PBE) form~\cite{33perdew1992atoms}.
A cutoff energy of 500 eV and vdW correction using the DFT-D3 method were applied~\cite{34johnson2005post}. 
The Brillouin zone was discretized using a $7\times7\times1$ Monkhorst-Pack $k$-grid~\cite{35monkhorst1976special}.
For geometry optimization, the convergence criteria for electron energy and force were set to $10^{-5}$ eV and 0.02 eV/{\AA}, respectively.
We treated the localized $d$ orbitals of the Mn atom using the GGA+$U$ method~\cite{liechtenstein1995density}, with an on-site Coulomb interaction $U$$_\text{eff}$ of 5.34 eV~\cite{0yan2021barrier}.
Ferroelectric polarization was assessed using the Berry phase method~\cite{36king1993theory}.
A vacuum region of 20 {\AA} was included in the structural relaxation calculations to prevent interactions between adjacent slabs.

The spin-polarized quantum transport properties were calculated using non-equilibrium Green’s functions combined with density functional theory~\cite{37taylor2001ab}, implemented in the Nanodcal software~\cite{38taylor2001ab}. The generalized gradient approximation (GGA) with the Perdew-Burke-Ernzerhof (PBE) functional was used for the electronic exchange-correlation calculations. In the electronic self-consistent calculations, we set the cutoff energy to 80 Hartree and applied a convergence criterion of \(10^{-4}\) eV for the Hamiltonian matrix. The temperature of the Fermi function was maintained at 300 K. To compute the current and electronic transmission coefficients, a $k$-point grid of \(100 \times 100 \times 1\) was utilized.

\begin{figure}[b]
	\centering
	\includegraphics[width=8.6cm,angle=0]{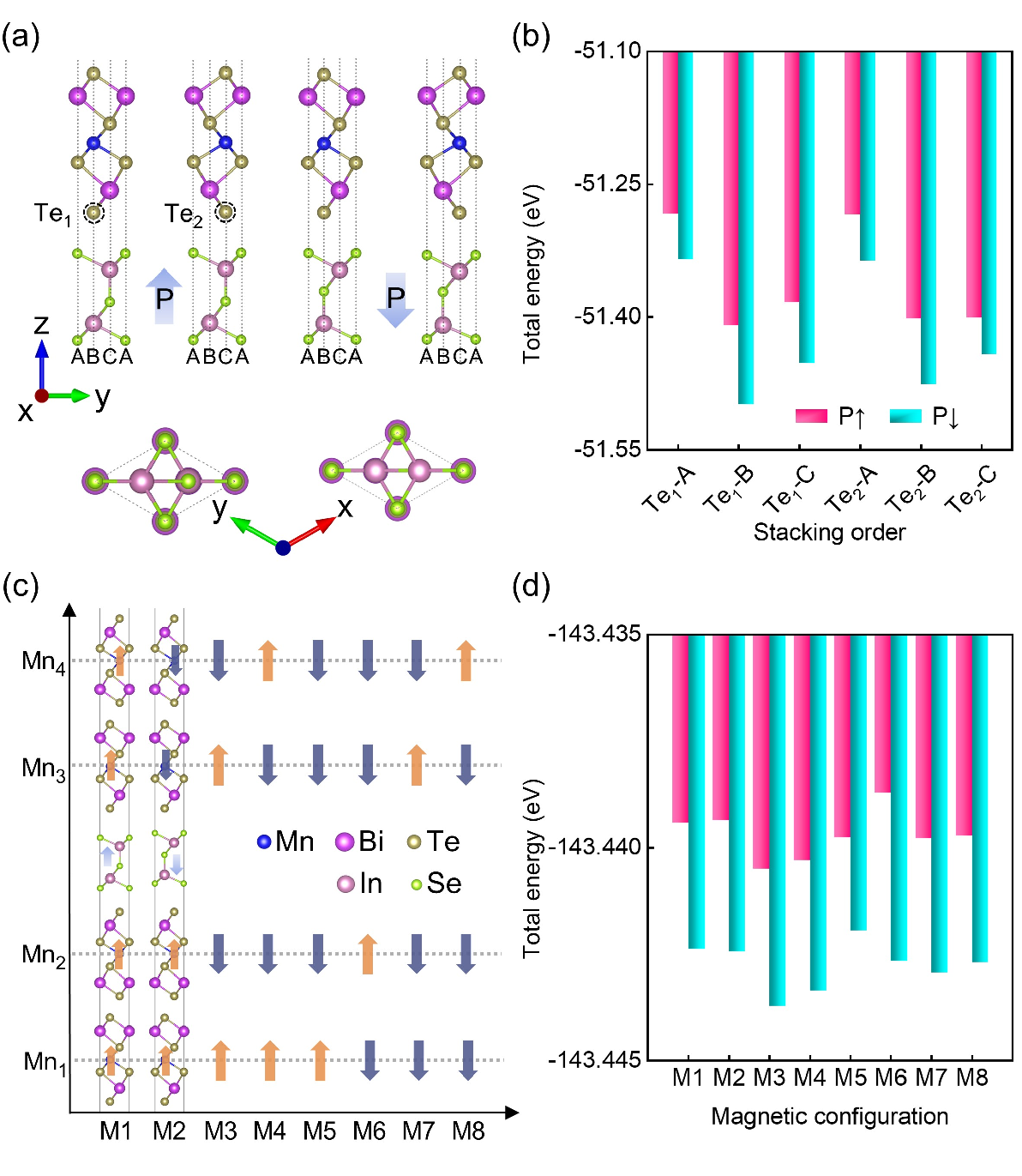}
	\caption{(a) Crystal structures of In$_2$Se$_3$/MnBi$_2$Te$_4$ heterostructure with two different ferroelectric polarizations. (b) The total energy of In$_2$Se$_3$/MnBi$_2$Te$_4$ in two polarized states as a function of different stacking orders. (c) Schematic of the magnetic setup of Mn atoms in MBT-2L/IS/MBT-2L. (d) The total energy of MBT-2L/IS/MBT-2L in both polarization states is a function of different magnetic configurations.}
	\label{Fig1}
\end{figure}

\begin{figure*}[htb!]
	\centering
	\includegraphics[width=18cm,angle=0]{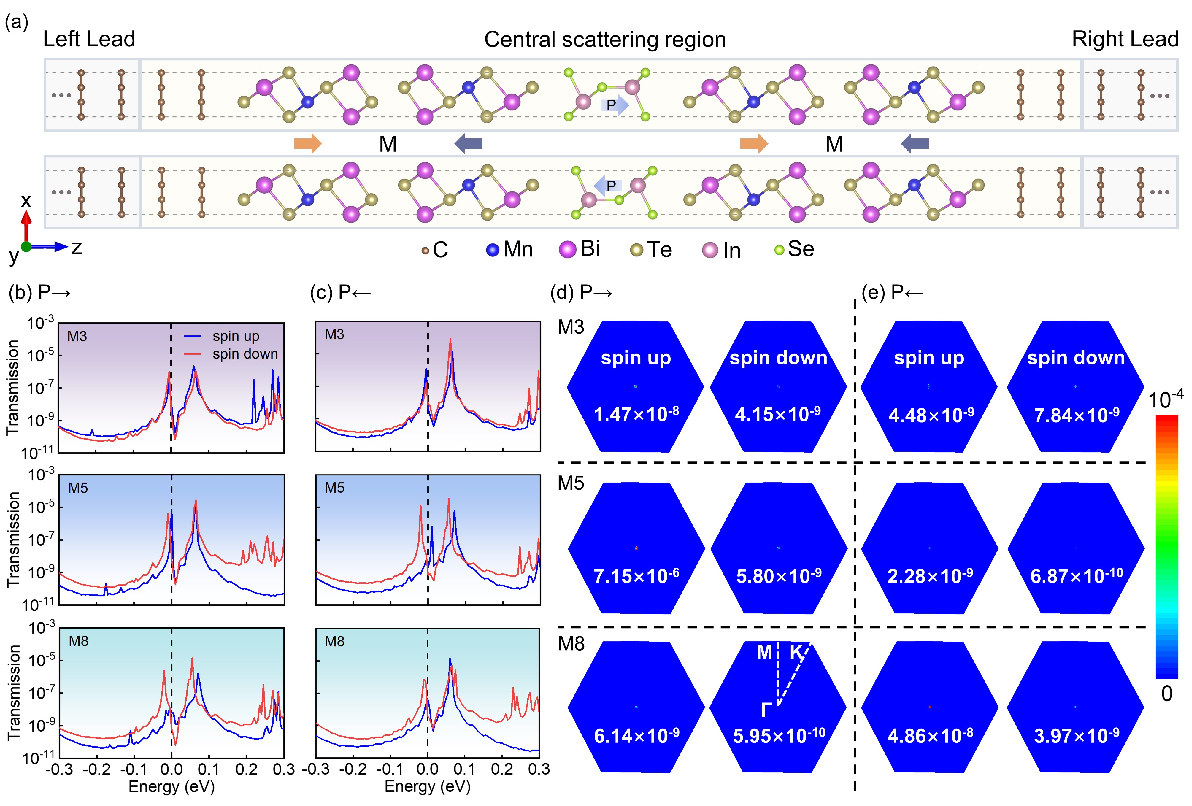}
	\caption{(a) The structure schematic diagrams of the MBT-2L/IS/MBT-2L AMFTJs with ferroelectric polarized in opposite directions. The spin-dependent electron transmission coefficients as a function of the Fermi energy at equilibrium for the AMFTJs with (b) P $\rightarrow$ and (c) P $\leftarrow$. The $k_\Arrowvert$-resolved electron transmission coefficients of the AMFTJs in the two-dimensional Brillouin zone at equilibrium with (d) P $\rightarrow$ and (e) P $\leftarrow$.}
	\label{Fig2}
\end{figure*}

The Landauer-Büttiker formula~\cite{39datta1997electronic,40meir1992landauer} is used to calculate the spin-polarized current $I_\sigma$ and conductance $G_\sigma$. The corresponding formulas are as follows:
\begin{align}
	I_\sigma &= \dfrac{e}{h} \int T_\sigma(E)[f_\text{L}(E) - f_\text{R}(E)] \mathrm{d}E,
\end{align}

\begin{align}
	G_\sigma &= \dfrac{e^{2}}{h}T_\sigma
\end{align}
Here, $\sigma$ indicates the spin orientation (either spin-up $\uparrow$ or spin-down $\downarrow$), $e$ represents the charge of an electron, $h$ denotes Planck’s constant, $T_\sigma(E)$ stands for the spin-specific transmission coefficient, and $f_\text{L(R)}(E)$ refers to the Fermi-Dirac distribution function of the left (right) lead.
The spin polarization rate ($\eta$) is expressed as follows:
\begin{align}
	\rm \eta &=\left|\dfrac{I_\uparrow - I_\downarrow}{I_\uparrow + I_\downarrow}\right|.
\end{align}
The TMR of MBT-2L/IS/MBT-2L AMFTJs at equilibrium can be defined as~\cite{0yang2024multistate}:
\begin{align}
	\rm TMR &=\frac{G_\text{M$X$}-G_\text{M3}}{G_\text{M3}}=\frac{T_\text{M$X$}-T_\text{M3}}{T_\text{M3}},
\end{align}
at bias voltage $V$, 
\begin{align}
	 \text{TMR}_{(V)} &=\frac{I_\text{M$X$}-I_\text{M3}}{I_\text{M3}},
\end{align}
here, M$X$ represents different magnetic configurations of the MBT-2L/IS/MBT-2L heterojunction, with M3 being the magnetic ground state (antiparallel configuration). $T_\text{M$X$/M3}$ and $I_\text{M$X$/M3}$ denote the total transmission coefficient at the Fermi level and the total current at bias voltage $V$ for magnetic state M$X$/magnetic ground state M3, respectively.
Another significant physical parameter, TER, can be defined as follows~\cite{42kang2020giant,43tao2016ferroelectricity}:
\begin{align}
	\rm TER &=\frac{|G_\uparrow - G_\downarrow|}{\text{min}(G_\uparrow, G_\downarrow)}=\frac{|T_\uparrow - T_\downarrow|}{\text{min}(T_\uparrow, T_\downarrow)},
\end{align}
at bias voltage $V$, 
\begin{align}
	 \text{TER}_{(V)} &=\frac{|I_\uparrow - I_\downarrow|}{\text{min}(I_\uparrow, I_\downarrow)},
\end{align}
where $T_{\uparrow/\downarrow}$ and $I_{\uparrow/\downarrow}$ represent the total transmission coefficient at the Fermi level and currents under a bias voltage $V$, which can be obtained by reversing the direction of the ferroelectric polarization of the barrier layer.

\begin{table*}[htb]
  \centering
  \renewcommand{\arraystretch}{1.5}
  \caption{Calculated spin-dependent electron transmission $T_{\uparrow}$ and $T_{\downarrow}$, TMR, TER, and spin polarization rate ($\eta$) for MBT-2L/IS/MBT-2L vdW AMFTJs at equilibrium.}
  \label{table1}
  \resizebox{\linewidth}{!}
  {
    \begin{threeparttable}
    \begin{tabular}{cccccccccccc}
    \toprule
    \multicolumn{2}{p{8.08em}}{Configuration} & \multicolumn{4}{c}{P $\rightarrow$} &       & \multicolumn{4}{c}{P $\leftarrow$} &  \\
\cmidrule{3-6}\cmidrule{8-11}    \multicolumn{2}{p{8.08em}}{and ratio} & \multicolumn{1}{c}{$T_{\uparrow}$} & \multicolumn{1}{c}{$T_{\downarrow}$} & \multicolumn{1}{c}{$T_\text{tot}=T_{\uparrow}+T_{\downarrow}$} & \multicolumn{1}{c}{\textit{$\eta$}} &       & \multicolumn{1}{c}{$T_{\uparrow}$} & \multicolumn{1}{c}{$T_{\downarrow}$} & \multicolumn{1}{c}{$T_\text{tot}=T_{\uparrow}+T_{\downarrow}$} & \multicolumn{1}{c}{\textit{$\eta$}} & \multicolumn{1}{c}{TER} \\
    \midrule
    \multicolumn{2}{c}{M1} & \multicolumn{1}{c}{$6.32 \times 10^{-10}$} & \multicolumn{1}{c}{$3.88 \times 10^{-9}$} & \multicolumn{1}{c}{$4.51 \times 10^{-9}$} & $72\%$  &       & \multicolumn{1}{c}{$6.25 \times 10^{-10}$} & \multicolumn{1}{c}{$2.29 \times 10^{-9}$} & \multicolumn{1}{c}{$2.91 \times 10^{-9}$} & $57\%$  & $55\%$ \\
    \multicolumn{2}{c}{M2} & \multicolumn{1}{c}{$2.88 \times 10^{-10}$} & \multicolumn{1}{c}{$4.35 \times 10^{-7}$} & \multicolumn{1}{c}{$4.35 \times 10^{-7}$} & $100\%$ &       & \multicolumn{1}{c}{$3.33 \times 10^{-8}$} & \multicolumn{1}{c}{$3.50 \times 10^{-10}$} & \multicolumn{1}{c}{$3.36 \times 10^{-8}$} & $98\%$  & $1194\%$ \\
    \multicolumn{2}{c}{M3} & \multicolumn{1}{c}{$1.47 \times 10^{-8}$} & \multicolumn{1}{c}{$4.15 \times 10^{-9}$} & \multicolumn{1}{c}{$1.88 \times 10^{-8}$} & $56\%$  &       & \multicolumn{1}{c}{$4.48 \times 10^{-9}$} & \multicolumn{1}{c}{$7.84 \times 10^{-9}$} & \multicolumn{1}{c}{$1.23 \times 10^{-8}$} & $27\%$  & $53\%$ \\
    \multicolumn{2}{c}{M4} & \multicolumn{1}{c}{$3.53 \times 10^{-7}$} & \multicolumn{1}{c}{$4.36 \times 10^{-9}$} & \multicolumn{1}{c}{$3.57 \times 10^{-7}$} & $98\%$  &       & \multicolumn{1}{c}{$1.90 \times 10^{-8}$} & \multicolumn{1}{c}{$4.40 \times 10^{-9}$} & \multicolumn{1}{c}{$2.35 \times 10^{-8}$} & $62\%$  & $1423\%$ \\
    \multicolumn{2}{c}{M5} & \multicolumn{1}{c}{$7.15 \times 10^{-6}$} & \multicolumn{1}{c}{$5.80 \times 10^{-9}$} & \multicolumn{1}{c}{$7.15 \times 10^{-6}$} & $100\%$ &       & \multicolumn{1}{c}{$2.28 \times 10^{-9}$} & \multicolumn{1}{c}{$6.87 \times 10^{-10}$} & \multicolumn{1}{c}{$2.97 \times 10^{-9}$} & $54\%$  & $241091\%$ \\
    \multicolumn{2}{c}{M6} & \multicolumn{1}{c}{$6.02 \times 10^{-9}$} & \multicolumn{1}{c}{$1.04 \times 10^{-8}$} & \multicolumn{1}{c}{$1.65 \times 10^{-8}$} & $27\%$  &       & \multicolumn{1}{c}{$1.26 \times 10^{-8}$} & \multicolumn{1}{c}{$1.85 \times 10^{-10}$} & \multicolumn{1}{c}{$1.27 \times 10^{-8}$} & $97\%$  & $29\%$ \\
    \multicolumn{2}{c}{M7} & \multicolumn{1}{c}{$1.13 \times 10^{-7}$} & \multicolumn{1}{c}{$2.82 \times 10^{-10}$} & \multicolumn{1}{c}{$1.14 \times 10^{-7}$} & $100\%$ &       & \multicolumn{1}{c}{$4.29 \times 10^{-9}$} & \multicolumn{1}{c}{$7.03 \times 10^{-9}$} & \multicolumn{1}{c}{$1.13 \times 10^{-8}$} & $24\%$  & $904\%$ \\
    \multicolumn{2}{c}{M8} & \multicolumn{1}{c}{$6.14 \times 10^{-9}$} & \multicolumn{1}{c}{$5.95 \times 10^{-10}$} & \multicolumn{1}{c}{$6.73 \times 10^{-9}$} & $82\%$  &       & \multicolumn{1}{c}{$4.86 \times 10^{-8}$} & \multicolumn{1}{c}{$3.97 \times 10^{-9}$} & \multicolumn{1}{c}{$5.26 \times 10^{-8}$} & $85\%$  & $682\%$ \\
    \midrule
    \multicolumn{1}{c}{\multirow{7}[14]{*}{TMR}} & M1    & \multicolumn{4}{c}{$-76\%$}     &       & \multicolumn{4}{c}{$-76\%$}     &  \\
\cmidrule{3-6}\cmidrule{8-11}          & M2    & \multicolumn{4}{c}{$2210\%$}    &       & \multicolumn{4}{c}{$173\%$}     &  \\
\cmidrule{3-6}\cmidrule{8-11}          & M4    & \multicolumn{4}{c}{$1796\%$}    &       & \multicolumn{4}{c}{$90\%$}      &  \\
\cmidrule{3-6}\cmidrule{8-11}          & M5    & \multicolumn{4}{c}{$37877\%$}   &       & \multicolumn{4}{c}{$-76\%$}     &  \\
\cmidrule{3-6}\cmidrule{8-11}          & M6    & \multicolumn{4}{c}{$-13\%$}     &       & \multicolumn{4}{c}{$3\%$}       &  \\
\cmidrule{3-6}\cmidrule{8-11}          & M7    & \multicolumn{4}{c}{$503\%$}     &       & \multicolumn{4}{c}{$-8\%$}      &  \\
\cmidrule{3-6}\cmidrule{8-11}          & M8    & \multicolumn{4}{c}{$-64\%$}     &       & \multicolumn{4}{c}{$327\%$}     &  \\
    \bottomrule
    \end{tabular}%
    \end{threeparttable}
  }
  \label{tab:addlabel}%
\end{table*}%

\section{Results and discussion}
\subsection*{A. The geometric structure model of AMFTJ}
The primary objective of this work is to design van der Waals antiferromagnetic-multiferroic tunnel junctions (vdW AMFTJs), which necessitates the identification of suitable materials for device fabrication. The first step in this process involves the exploration for vdW antiferromagnetic materials.
Layered vdW MnBi$_2$Te$_4$ has been successfully synthesized experimentally and confirmed to exhibit intriguing interlayer antiferromagnetism and intralayer ferromagnetism~\cite{li2019intrinsic,gong2019experimental}, which has immediately attracted attention in the field of spintronics. Recent experimental~\cite{zhang2019experimental,chen2024even} and theoretical studies~\cite{li2022identifying,zhan2022spin,0yan2021barrier,0dong2023spin} have demonstrated the feasibility of fabricating spintronic devices based on MnBi$_2$Te$_4$ and investigating their transport properties. Therefore, MnBi$_2$Te$_4$ is a promising candidate material for vdW AMFTJs.
The crystal structure of MnBi$_2$Te$_4$ consists of Te-Bi-Te-Mn-Te-Bi-Te septuple layers stacked along the $c$-axis through the vdW interaction.
We chose a typical In$_2$Se$_3$ monolayer~\cite{46ding2017prediction} as the ferroelectric barrier for AMFTJs, which consists of five atomic layers arranged in the sequence Se-In-Se-In-Se.
To verify the ferroelectricity of In$_2$Se$_3$, we calculated the ferroelectric polarization to be 0.095 $e$·{\AA}/unit cell based on the Berry phase method~\cite{42kang2020giant}, which is in agreement with a previous study~\cite{46ding2017prediction}.
The crystal structures of In$_2$Se$_3$/MnBi$_2$Te$_4$ heterostructures with two different ferroelectric polarizations are depicted in Fig.~\ref{Fig1}(a). 
To determine the stacking configuration of the ferroelectric barrier layer In\(_2\)Se\(_3\) with MnBi\(_2\)Te\(_4\) on either side, we design four different configurations of the In\(_2\)Se\(_3\)/MnBi\(_2\)Te\(_4\) interface due to the two opposite ferroelectric polarization directions. Additionally, monolayer In\(_2\)Se\(_3\) harbors three different sites (A/B/C) along the in-plane direction [see Fig.~\ref{Fig1}(a)]. Therefore, each polarization direction of the In\(_2\)Se\(_3\)/MnBi\(_2\)Te\(_4\) interface has six different high-symmetry stacking configurations, i.e., Te\(_{1/2}\)-A/B/C. As shown in Fig.~\ref{Fig1}(b), by calculating the total energy, it is found that the energy-favored stacking configuration for both polarization directions is Te\(_1\)-B.
We then construct the core composition of the central region of the AMFTJ device, which consists of bilayer MnBi\(_2\)Te\(_4\)/In\(_2\)Se\(_3\)/bilayer MnBi\(_2\)Te\(_4\) (MBT-2L/IS/MBT-2L). It is noteworthy that the bilayer MnBi\(_2\)Te\(_4\) maintains its bulk interlayer stacking configuration.
As displayed in Fig.~\ref{Fig1}(c), for the MBT-2L/IS/MBT-2L heterojunction, there are eight possible magnetic orderings (refer to the arrows labeled M1-8 for detailed configurations). 
The electronic band structures of MBT-2L/IS/MBT-2L in these resistance states are summarized in Fig.~\ref{Fig7} in Appendix A. A detailed analysis of the electronic structure is presented in Appendix A. The main result is that reversing the ferroelectric polarization direction of In\(_2\)Se\(_3\) under the same magnetic arrangement switches the spin channel, indicating a robust magnetoelectric coupling effect. This contributes to intriguing transport phenomena in AMFTJ devices.
To verify that the ferroelectricity of In\(_2\)Se\(_3\) is not disrupted when fabricated into an AMFTJ, we calculate the electrostatic potential along the \( z \)-direction of the MBT-2L/IS/MBT-2L heterojunction under two opposite ferroelectric polarizations. The results are presented in Fig.~\ref{Fig8} in Appendix A. 
One can observe that the reversal of the ferroelectric polarization results in the exchange of peak positions in the electrostatic potential (as indicated by the diagonal lines in Fig.~\ref{Fig8}), implying that the ferroelectricity of In\(_2\)Se\(_3\) is preserved and holds potential for fostering TER effects.
We also calculate the energies of these eight magnetic states and find that the magnetic ground state for both polarization states is M3, i.e., interlayer antiferromagnetism as shown in Fig.~\ref{Fig1}(d).

After determining the optimal stacking sequence, ferroelectricity and magnetic ground state of the MnBi\(_2\)Te\(_4\)/In\(_2\)Se\(_3\) heterojunction, we can now construct the AMFTJs device. As shown in Fig.~\ref{Fig2}(a), graphite is used as the semi-infinite electrode, with two layers of graphite serving as a buffer layer. The presence of the buffer layer prevents wave function coupling between the MnBi\(_2\)Te\(_4\) in the central region and the graphite electrodes.
At the graphene/MnBi\(_2\)Te\(_4\) interface, there are two high-symmetry stacking configurations: Te-C and Te-hollow. The energetically favorable structure is the Te-hollow configuration (see Fig.~\ref{Fig9} in Appendix B).
The optimized in-plane lattice constants of monolayer MnBi$_2$Te$_4$, In$_2$Se$_3$, and graphene are 4.336 {\AA}~\cite{mbt-otrokov2019unique}, 4.106 {\AA}~\cite{42kang2020giant}, and 2.460 {\AA}~\cite{gr-yuan2024tunneling}, respectively. 
Therefore, the minimum unit matches for constructing AMFTJs are $1\times1$, $1\times1$, and $\sqrt{3}\times\sqrt{3}$, respectively. 
Considering MnBi\(_2\)Te\(_4\) as the majority layer in the central region, we use its in-plane lattice constant as the lattice parameter for the entire central region. In this case, the mismatch with graphite/In\(_2\)Se\(_3\) is 1.73\%/5.30\%.
These mismatch rates are unlikely to be observed experimentally due to the weak vdW interaction between the layers in the vdW system~\cite{48li2019spin}. However, due to the periodic boundary conditions, they have to be considered in theoretical simulations. In the subsequent part of this work, we also investigate the effect of biaxial strain on the transport properties.

\subsection*{B. Electron transport properties and strain effects in equilibrium}
As mentioned above, the multilayer MnBi$_2$Te$_4$ exhibits 8 magnetic configurations (M1-8), while the ferroelectric In$_2$Se$_3$ has two ferroelectric polarization states (P $\rightarrow$/P $\leftarrow$). This implies that 16 possible combined states can be induced in the AMFTJ. We first investigate the TMR and TER effects of the vdW AMFTJ at equilibrium.
As summarized in \autoref{table1}, TMR/TER depends on different magnetic configurations and the ferroelectric polarization direction of In$_2$Se$_3$.
It is worth noting that the tiny value of $T_{\uparrow/\downarrow}$ is converged by our $k$-mesh test.
In the right ferroelectric polarization (P $\rightarrow$), the highest TMR is achieved in the M5 configuration, reaching 37877\%. In the left ferroelectric polarization (P $\leftarrow$), the highest TMR is in the M8 configuration, at 327\%. Besides the TMR effect, the TER effect is also crucial for evaluating the transport performance of AMFTJ. As shown in \autoref{table1}, the highest TER ratio of 241091\% occurs in the M5 configuration with P $\rightarrow$. Such high TMR/TER ratios are sufficient for applications in spintronic devices, such as magnetic sensors, hard disk read heads, and magnetoresistive random-access memory.
We also observe perfect spin filtering effects ($\eta$=100\%), occurring in the M2/5/7 configurations with P $\rightarrow$ polarization, suggesting that our AMFTJ has potential applications in spin filters.

\begin{figure}[htp!]
	\centering
	\includegraphics[width=8.6cm,angle=0]{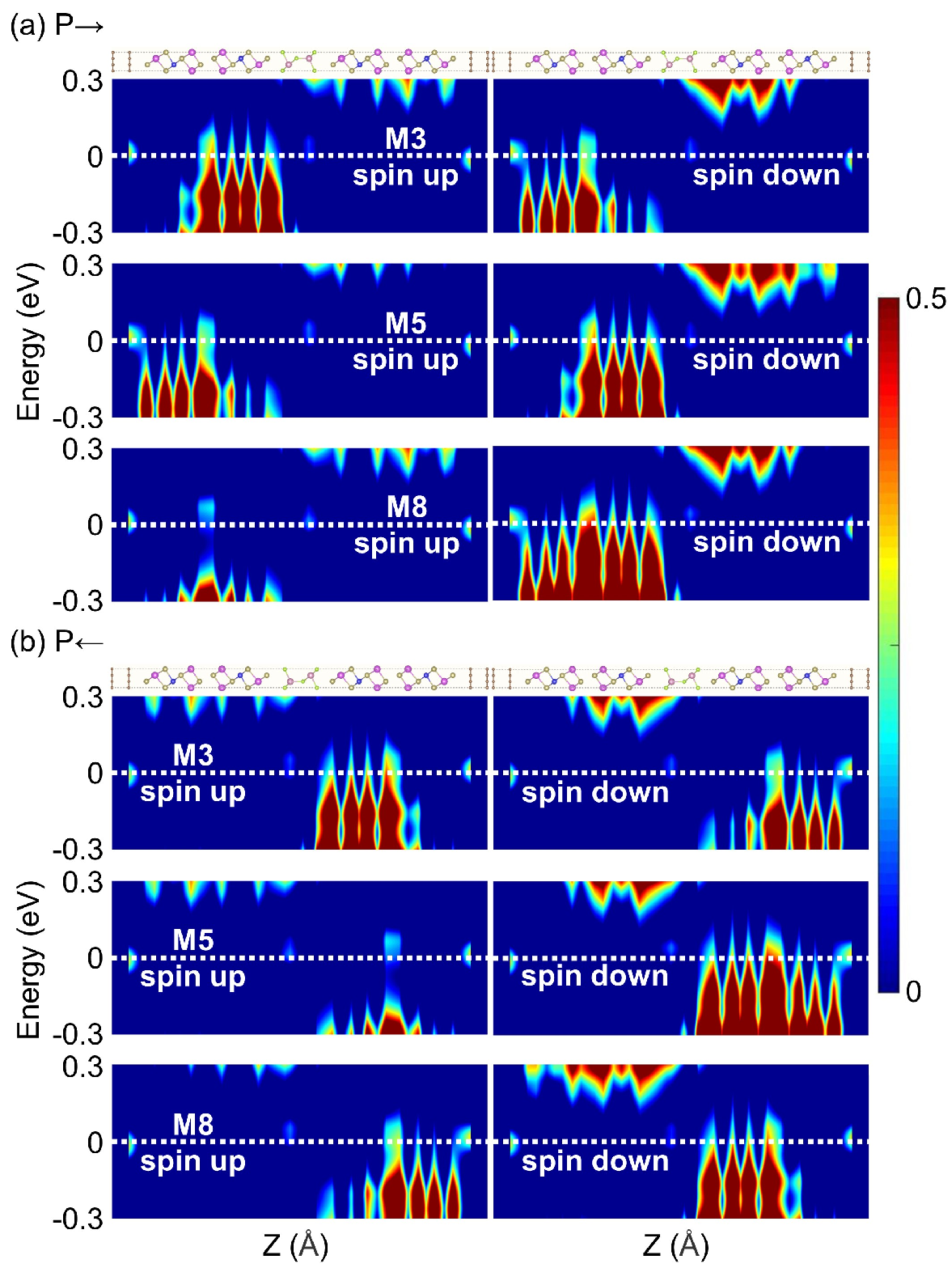}
	\caption{The averaged spin-resolved PDOS distribution over the $xy$ plane of the central scattering regions along the transport direction ($z$ axis) for MBT-2L/IS/MBT-2L AMFTJs in the equilibrium state. The white horizontal dashed line represents the Fermi level. (a) P $\rightarrow$; (b) P $\leftarrow$.}
	\label{Fig3}
\end{figure}

\begin{figure}[htp!]
	\centering
	\includegraphics[width=8.6cm,angle=0]{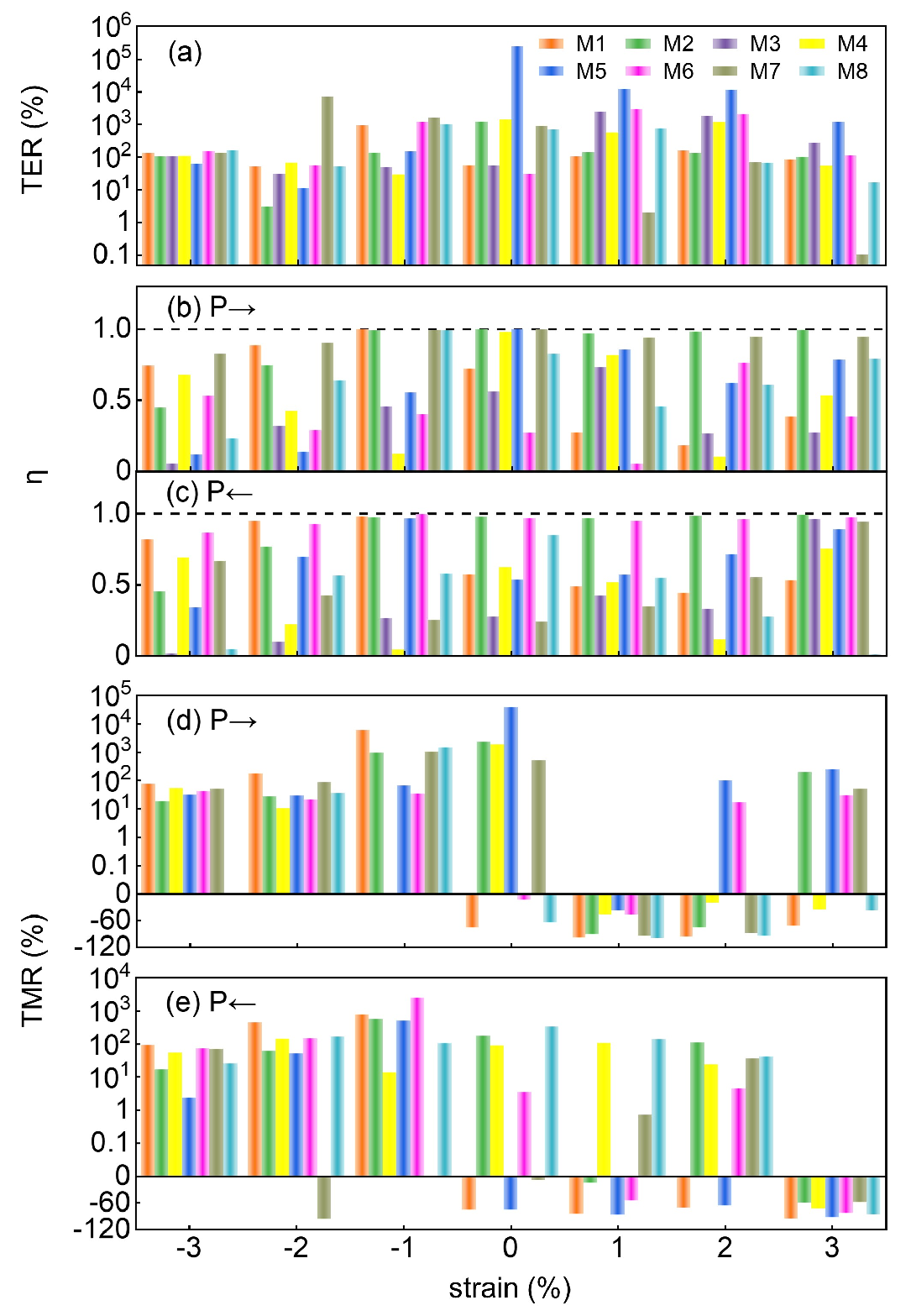}
	\caption{The (a) TER, (b), (c) spin polarization rate ($\eta$) and (d), (e) TMR of MBT-2L/IS/MBT-2L AMFTJs in different ferroelectric polarizations of in-plane biaxial strain.}
	\label{Fig4}
\end{figure}

Next, we calculate the electron transmission spectra within the energy range of -0.3 eV to 0.3 eV for these 16 resistance states at equilibrium. The results are presented in Figs.~\ref{Fig2}(b) and (c) and Fig.~\ref{Fig10} in Appendix C. Based on the maximum TMR/TER and the magnetic ground state, we use M3, M5, and M8 as examples for further discussion. The results for the other magnetic configurations are presented in the Appendix.
As shown in Fig.~\ref{Fig2}(b), M3 is the magnetic ground state, where the bilayers of MnBi$_2$Te$_4$ on both sides of the barrier layer are in an antiferromagnetic configuration, theoretically resulting in a net magnetic moment of zero. However, the electron transmission spectrum of M3 is not completely degenerate. This is mainly due to the lifting of band degeneracy by the ferroelectric polarization field. 
Compared to Fig.~\ref{Fig2}(c), a clear spin channel exchange is observed due to the reversal of the ferroelectric polarization direction.
For the M5 and M8 configurations, the net magnetic moment is approximately twice that of a single layer of MnBi$_2$Te$_4$. Therefore, significant spin polarization appears in the electron transmission spectrum, and larger spin splitting occurs at higher energies. This implies that a high spin filtering efficiency can be achieved by shifting the Fermi level.
The distribution of $k_\Arrowvert$-resolved electron transmission coefficients at the Fermi level within the two-dimensional Brillouin zone can be used to understand the large TMR and TER obtained at equilibrium.
From Figs.~\ref{Fig2}(d) and (e), it is evident that at the Fermi level, electrons are transmitted only at the $\Gamma$ point. This is mainly because graphite, used as an electrode, has conductive channels only at the Dirac point. Additionally, the in-plane supercell of graphite is a special $\sqrt{3} \times \sqrt{3}$ configuration, which folds the Dirac point from the K point to the $\Gamma$ point.
For M5, 'hot spots' appear only in the P $\rightarrow$ ferroelectric polarization state and not in the P$\leftarrow$ state, implying a large TER. Additionally, the spin-up transmission is significantly greater than the spin-down transmission, indicating an almost perfect spin-filtering effect.
Compared to M3, M5 exhibits significantly more 'hot spots' in the P$\rightarrow$ state, leading to a large TMR ratio obtained from Eq. (4).
The distribution of transmission coefficients at the Fermi level for the remaining magnetic configurations in the 2D Brillouin zone is shown in Fig.~\ref{Fig11} in Appendix C. Similar analyses can be directly applied, and therefore will not be repeated here.

\begin{figure*}[htbp]
	\centering
	\includegraphics[width=18cm,angle=0]{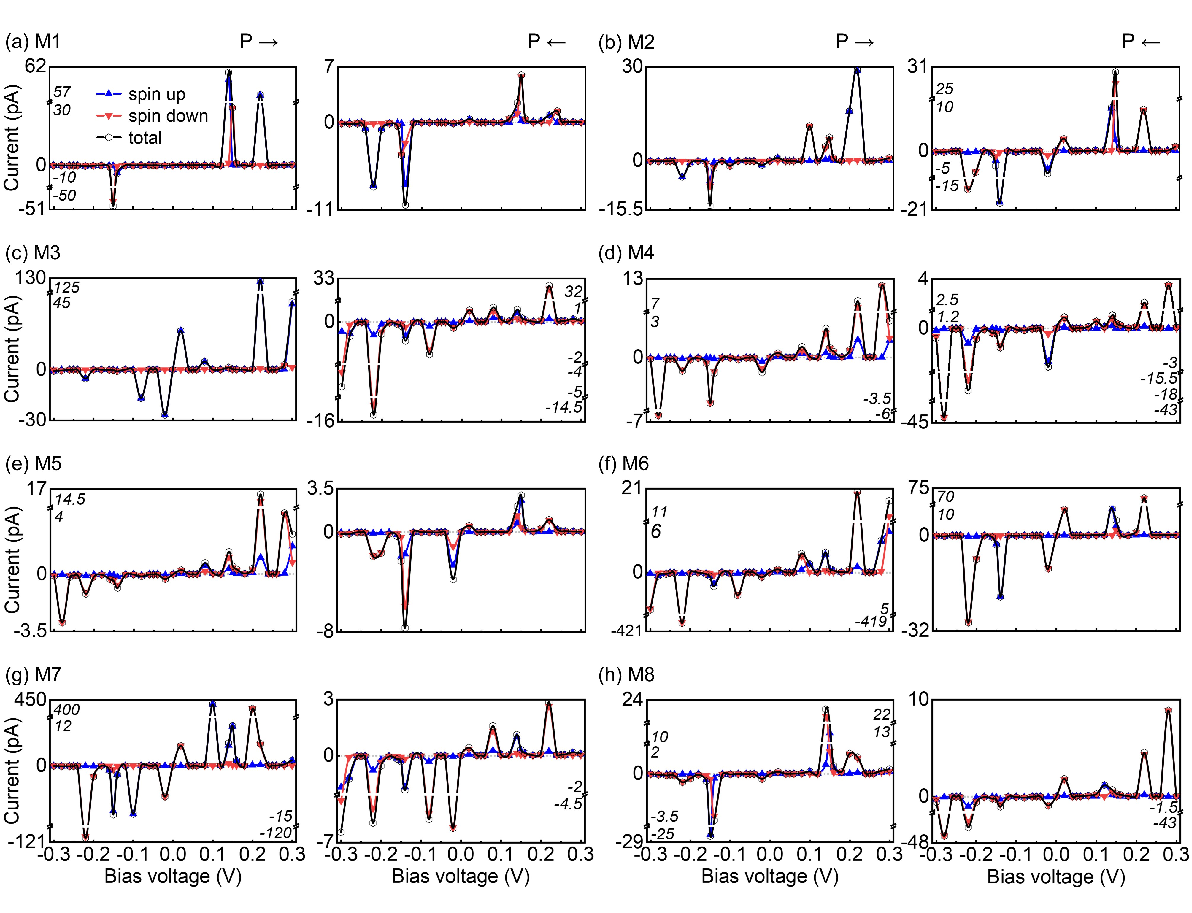}
	\caption{The variation of the spin current (a)-(h) $vs$ the bias voltages for MBT-2L/IS/MBT-2L AMFTJ in different ferroelectric polarizations.}
	\label{Fig5}
\end{figure*}

\begin{figure*}[htb!]
	\centering
	\includegraphics[width=18cm,angle=0]{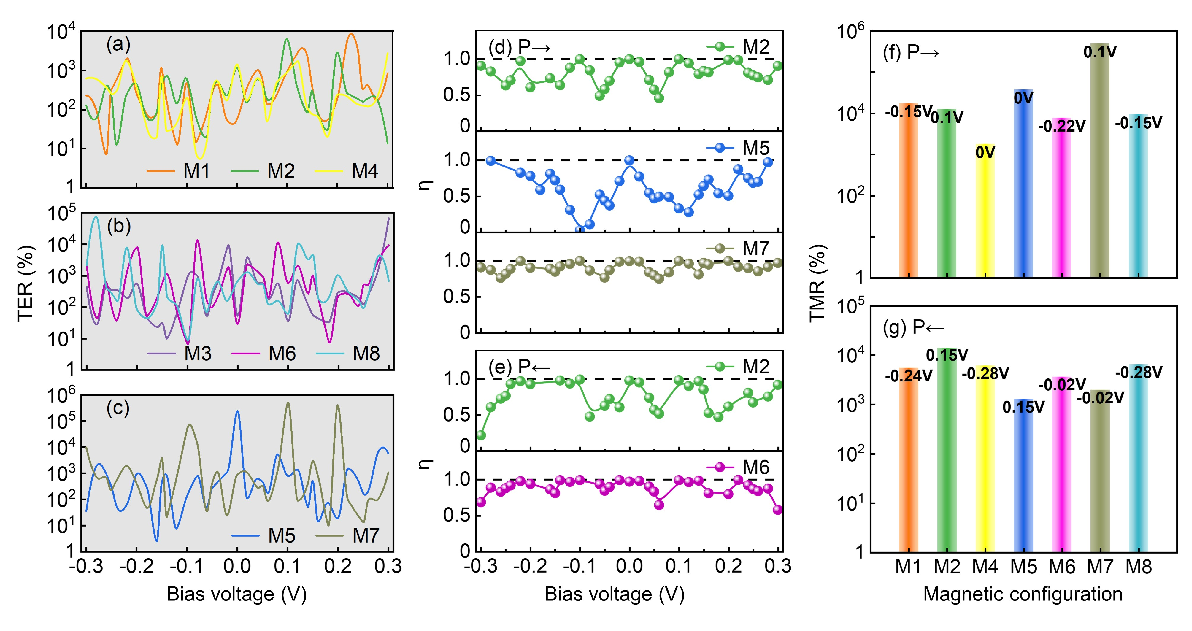}
	\caption{The variation of the TER ratios (a), (b) and (c), spin polarization rate ($\eta$) (d) and (e), and TMR ratios (f) and (g) $vs$ the bias voltages for MBT-2L/IS/MBT-2L vdW AMFTJs in different ferroelectric polarizations.}
	\label{Fig6}
\end{figure*}

To provide a clearer illustration of the significant TMR and TER effect, we calculate the real-space atomic projected density of states (PDOS) along the transport direction ($z$-axis) in the central scattering region at equilibrium for all magnetic order, as shown in Fig.~\ref{Fig3} and Fig.~\ref{Fig12} in Appendix C.
One can observe that the low electronic density of states (DOS) in the In\(_2\)Se\(_3\) region for all the PDOS plots indicates that the In\(_2\)Se\(_3\) serves as a sufficiently thick insulating barrier layer, and that electron transport is primarily dominated by the bilayer MnBi\(_2\)Te\(_4\).
In detail, for the M3 configuration in the P \(\rightarrow\) state [see Fig.~\ref{Fig3}(a)], along the positive \(z\)-axis transport direction, spin-up and spin-down electrons at the Fermi level flow out from the MnBi\(_2\)Te\(_4\) bilayer with majority states, tunnel through the In\(_2\)Se\(_3\) barrier layer, and finally pass through the MnBi\(_2\)Te\(_4\) bilayer with minority states on the right side. This correspondence between majority states and minority states indicates the high-resistance electron transport behavior.
Although the M5 configuration in the P\(\rightarrow\) state exhibits similar high-resistance electron transport characteristics as M3, more electrons participate in the transport due to the majority states of spin-up MnBi\(_2\)Te\(_4\) being closer to the left graphite electrode in M5. Additionally, the barrier layer thickness traversed by spin-down electrons in the P\(\rightarrow\) state for the M5 configuration is smaller than that for the corresponding M3 configuration, leading to a higher number of electrons passing through. These results are consistent with our calculated electron transmission coefficients at the Fermi level under equilibrium conditions, thus the M5 configuration in the P\(\rightarrow\) state exhibits a significant TMR effect. The magnetic configurations of M8 and the rest are similar to the above analysis and can be directly referenced.
Compared to the PDOS of the M5 configuration in the P\(\rightarrow\) state, it is evident that the M5 configuration in the P\(\leftarrow\) state has almost no DOS at the Fermi level for spin-up electron transport channels [see Fig.~\ref{Fig3}(b)], indicating an extremely high resistance state. For the spin-down channel, there is almost no DOS near the left graphite electrode, suggesting that even fewer electrons can pass through compared to the P\(\rightarrow\) state. Therefore, the significant difference in electron transport capabilities between these two ferroelectric polarization directions in the M5 configuration leads to a large TER effect.

Previous works have shown that applying in-plane biaxial strain is an effective measure to improve the transport properties of van der Waals tunnel junction devices~\cite{0yan2024giant,50yan2022giant}. Considering the mismatch at our AMFTJ interface, we systematically investigate the effect of in-plane biaxial strain in the range of -3\% to 3\% (in intervals of 1\%) on different magnetic configurations of the AMFTJ at equilibrium. The results are shown in Fig.~\ref{Fig4}.
It can be found that biaxial strain has a significant effect on the regulation of TER, TMR, and spin polarization rate ($\eta$).
For the TER effect, as shown in Fig.~\ref{Fig4}(a), one can observe that the TER generally decreases with increasing tensile strain for different magnetic states, while under compressive strain, it initially increases and then decreases.
Specifically, for instance, in the M7 configuration, the TER reaches its maximum at 2\% compressive strain, increasing by an order of magnitude compared to its value in the equilibrium state.
In the P→ state, as shown in Fig.~\ref{Fig4}(b), we observe that the $\eta$ of M1 and M8 first decreases and then increases with increasing tensile strain, whereas under compressive strain, it first increases and then decreases, approaching nearly perfect spin filtering at 1\%. For M2 and M7, the $\eta$ remains robust under tensile strain, maintaining almost 100\%, but gradually decreases with increasing compressive strain. The $\eta$ of M3 and M5 decreases with increasing compressive strain. For M4 and M5, the $\eta$ first decreases and then increases with increasing tensile strain. M6 exhibits an irregular oscillation effect.
In the P← state, as shown in Fig.~\ref{Fig4}(c), the $\eta$ of M1 and M5 first increases and then decreases with increasing compressive strain, approaching nearly 100\% at 1\%. The $\eta$ of M2 and M6 remains nearly 100\% under tensile strain. The $\eta$ of M3 and M5 increases with increasing tensile stress and first increases and then decreases with increasing compressive strain. The $\eta$ of M4 first decreases and then increases with increasing tensile or compressive strain. The $\eta$ of M7 (M8) increases (decreases) with increasing strain.
Unlike the TER effect, the TMR effect undergoes significant changes under strain. As shown in Figs.~\ref{Fig4}(b) and (e), applying stress can not only effectively modulate the magnitude of the TMR in MBT-2L/IS/MBT-2L AMFTJ but also interestingly change its sign. Specifically, in the P→ state, the TMR of M6 can change from a negative value to a positive value under tensile strain and increase by two orders of magnitude. Moreover, in the P→ state, the TMR of M1, M4, and M5 is positive under tensile strain and negative under compressive strain. A similar phenomenon is observed for the TMR of M1 and M5 in the P← state.
The effective modulation of the transport properties of AMFTJ devices by external strain is mainly attributed to the biaxial strain altering the in-plane and out-of-plane atomic lattice structures, ultimately leading to changes in the electronic properties.

\begin{figure*}[htb!]
	\centering
	\includegraphics[width=18cm,angle=0]{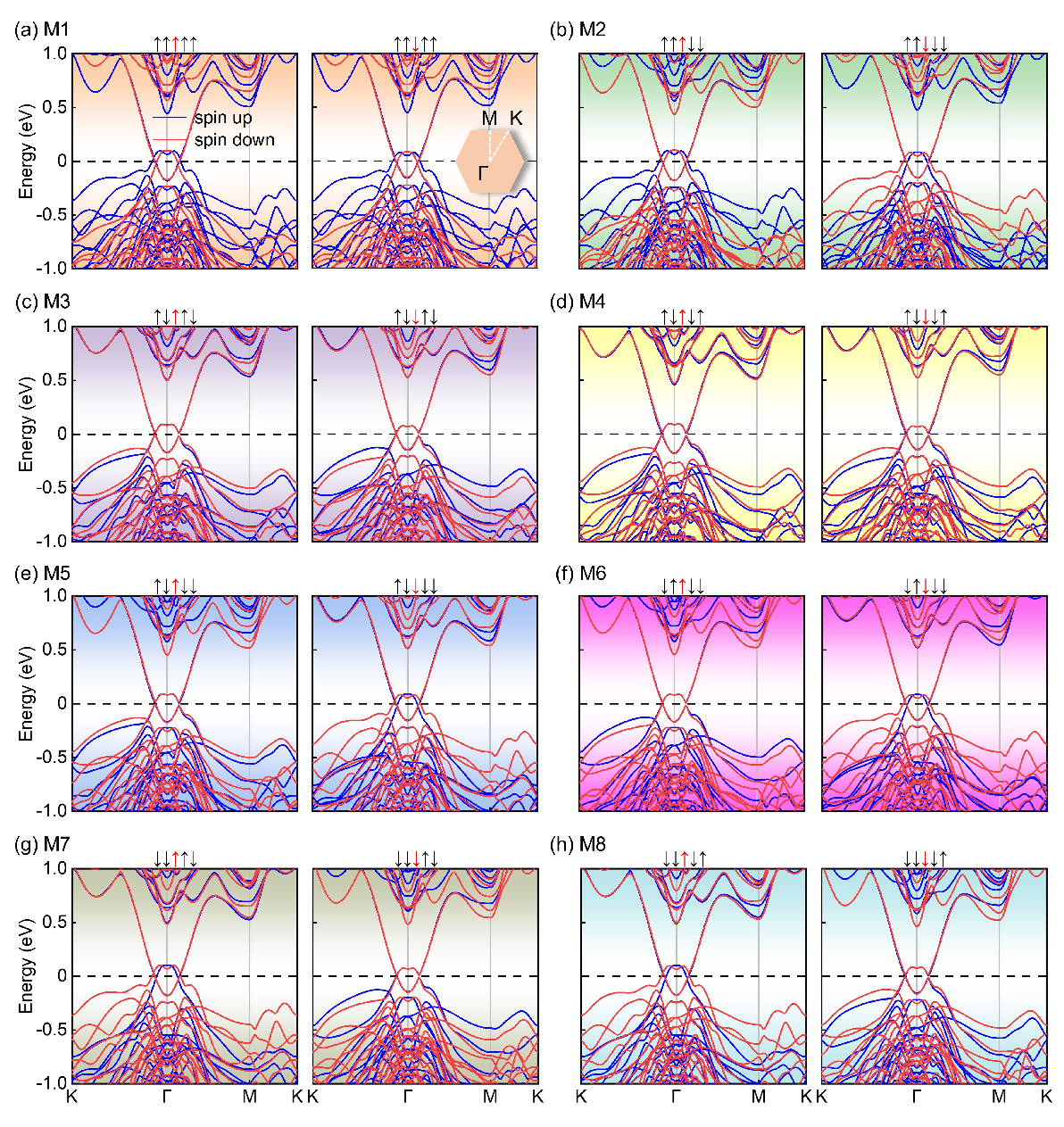}
	\caption{(a)-(h) The band structures of different magnetic configurations of MBT-2L/IS/MBT-2L heterostructure under different ferroelectric polarizations. The Fermi level is set to zero.}
	\label{Fig7}
\end{figure*}

\subsection*{C. Bias voltage-tunable electron transport properties}
The ultimate service goal of AMFTJ is to serve as a micro-unit for information storage integrated into spintronic devices. Therefore, the evaluation of its electronic transport properties in a non-equilibrium state is particularly important. In this section, as depicted in Fig.~\ref{Fig5} and Fig.~\ref{Fig6}, we calculate the spin current, TMR, TER, and $\eta$ of the MBT-2L/IS/MBT-2L vdW AMFTJ in the bias range from -0.3 V to 0.3 V under the P$\rightarrow$/P$\leftarrow$ state for various magnetic configurations.
Note that the bias voltage V$_b$ is set by applying the chemical potential on the left (right) electrode as $+V_b/2$ ($-V_b/2$).
Overall, as shown in Fig.~\ref{Fig5}, the I-V (current-voltage) curves of the MBT-2L/IS/MBT-2L AMFTJ exhibit differences under various magnetic configurations and ferroelectric polarization states. However, they share common features, specifically displaying oscillatory behavior and multiple peaks of negative differential resistance (NDR) effects.
Additionally, at certain specific bias voltages (around ±0.2 V), the I-V curves reveal perfect spin filtering effects, and the spin channels can be flexibly switched by different magnetic orderings and ferroelectric polarization directions [see Figs.~\ref{Fig5}(a),(c),(f), and (g)].

As illustrated in Figs.~\ref{Fig6}(a)-(c), the oscillatory I-V curves result in the TER of AMFTJ with different magnetic configurations also exhibiting oscillatory behavior under applied bias voltages, and significantly enhancing the TER ratio. This enhancement can reach up to two orders of magnitude at certain bias voltages. For example, the TER of M1, M3, and M6 at equilibrium are 55\%, 53\%, and 29\%, respectively, but the corresponding TMR values at bias voltages of 0.22 V (M1), 0.3 V (M3), and -0.08 V (M6) are $6\times10^{3}$\%, $6.82\times10^{4}$\%, and $1.36\times10^{4}$\%, respectively. Furthermore, the maximum TER at equilibrium is $2.4\times10^{5}$\%, which increases to a maximum of $4.97\times10^{5}$\% under bias voltages.
Based on the magnetic configurations and ferroelectric polarization states that achieve perfect spin filtering effects in the equilibrium state, i.e., M2, M5, and M7 in the P→ state, as well as M2 and M6 in the P← state, we calculate the evolution of their $\eta$ with bias voltage, as shown in Figs.~\ref{Fig6}(d) and (e).
One can observe that for M2 in the P→ state, the $\eta$ initially decreases and then increases with the applied bias voltage (including both positive and negative). It generally decreases from 100\% to around 50\% and then increases back to 100\%, maintaining perfect spin filtering effects at certain specific bias voltages. The evolution of $\eta$ with bias voltage for M5 is largely similar to that of M2, with the distinction that the $\eta$ of M5 decreases to a trough value of approximately 10\%. In contrast, for M7, the $\eta$ exhibits robustness against the applied bias voltage, consistently maintaining excellent spin filtering effects.
From Fig.~\ref{Fig6}(e), it can be clearly observed that in the P← state, the $\eta$ of M2 exhibits a regular oscillatory effect with the increase of positive bias voltage. Specifically, the $\eta$ first decreases then increases, decreases again, and increases again, with peak values around 100\% and trough values around 50\%. As the negative bias voltage increases, the $\eta$ first decreases then increases, and decreases again to around 20\%.
For M6 in the P← state, the $\eta$ remains consistently high with increasing bias voltage, except at the boundary values of ±0.3 V. Moreover, it achieves perfect spin filtering at most bias values.

In Fig.~\ref{Fig13} of Appendix D, we also calculated the bias voltage-dependent TMR effect of the MBT-2L/IS/MBT-2L AMFTJ. The maximum TMR at specific bias voltages for each magnetic configuration in different ferroelectric polarization states is presented in Figs.~\ref{Fig6}(f) and (g).
As shown in Fig.~\ref{Fig13} of Appendix D, applying a bias voltage can not only significantly increase the TMR values but also flexibly change their sign.
In the P→ state, the maximum TMR of $5.01\times10^{5}$\% occurs for magnetic configuration M7 when a 0.1 V bias voltage is applied [see Fig.~\ref{Fig6}(f)], whereas in the P← state, the maximum TMR of $1.40\times10^{4}$\% occurs for magnetic configuration M2 when a 0.15 V bias voltage is applied [see Fig.~\ref{Fig6}(g)].
In general, the maximum TMR obtained by applying a bias voltage to these different magnetic configurations of AMFTJs in different polarization states is on the order of $10^{3}$\% or higher. Such high values have great potential for applications in spintronic devices.

\begin{figure}[htb!]
	\centering
	\includegraphics[width=8.5cm,angle=0]{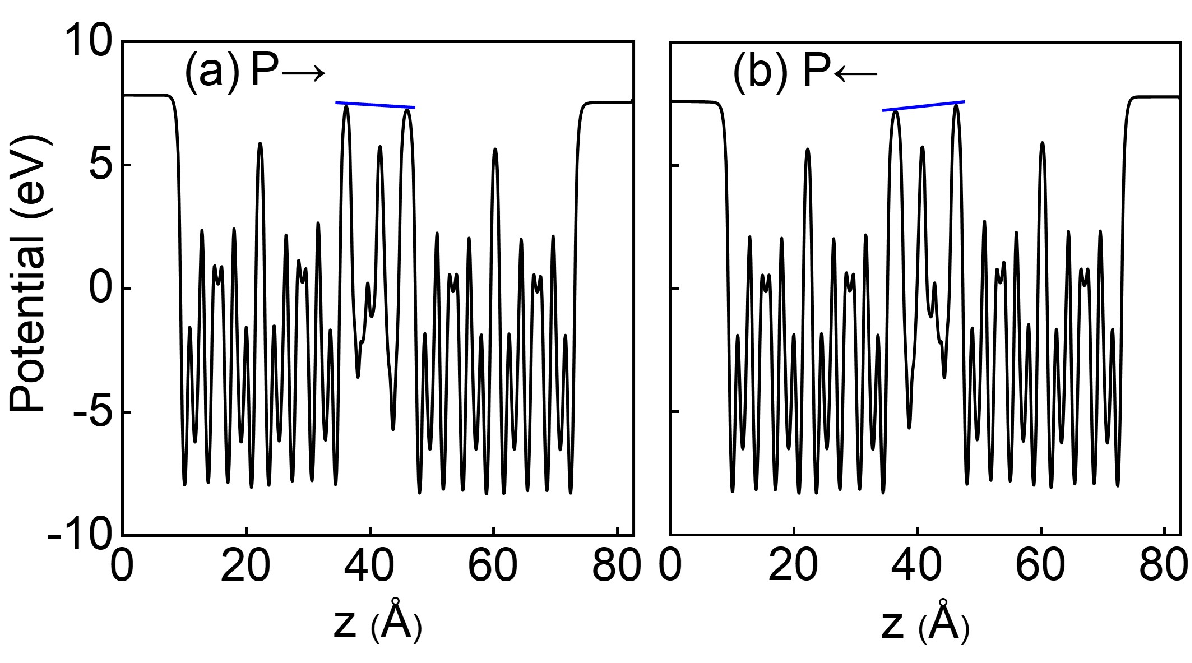}
	\caption{The evolution of the electrostatic potential of the MBT-2L/IS/MBT-2L heterostructure along the $z$ axis. (a) P $\rightarrow$; (b) P $\leftarrow$.}
	\label{Fig8}
\end{figure}

\begin{figure}[htb!]
	\centering
	\includegraphics[width=8.6cm,angle=0]{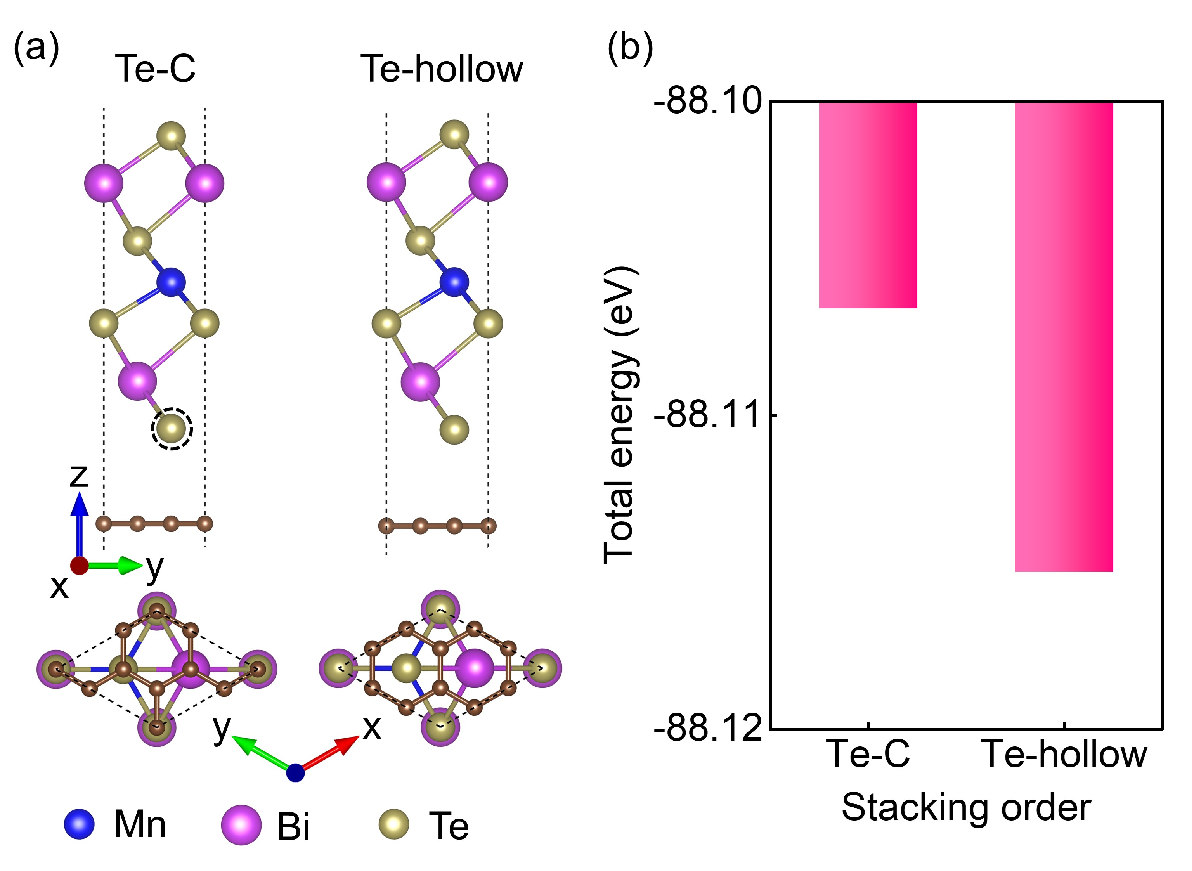}
	\caption{(a) Different stacking orders of MnBi$_2$Te$_4$/graphene heterojunction. (b) The evolution of total energy with different stacking orders.}
	\label{Fig9}
\end{figure}

\begin{figure*}[htbp]
	\centering
	\includegraphics[width=18cm,angle=0]{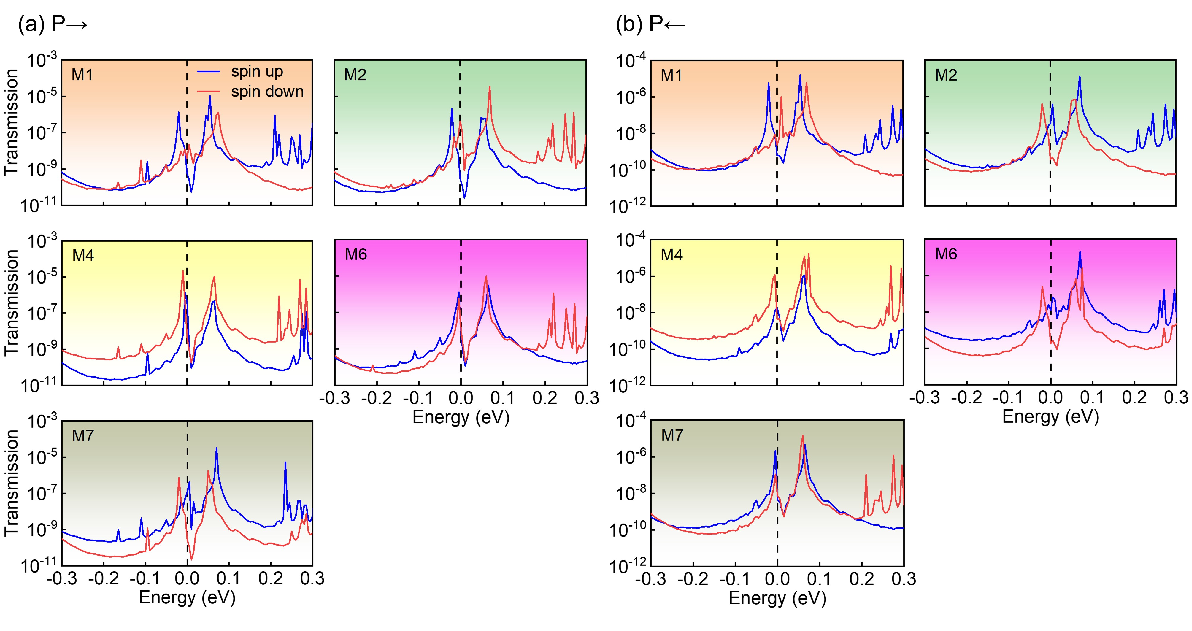}
	\caption{The spin-dependent zero-bias transmission coefficient curves of the MBT-2L/IS/MBT-2L AMFTJs with (b) P $\rightarrow$ and (c) P $\leftarrow$}
	\label{Fig10}
\end{figure*}

\begin{figure*}[htbp]
	\centering
	\includegraphics[width=17.5cm,angle=0]{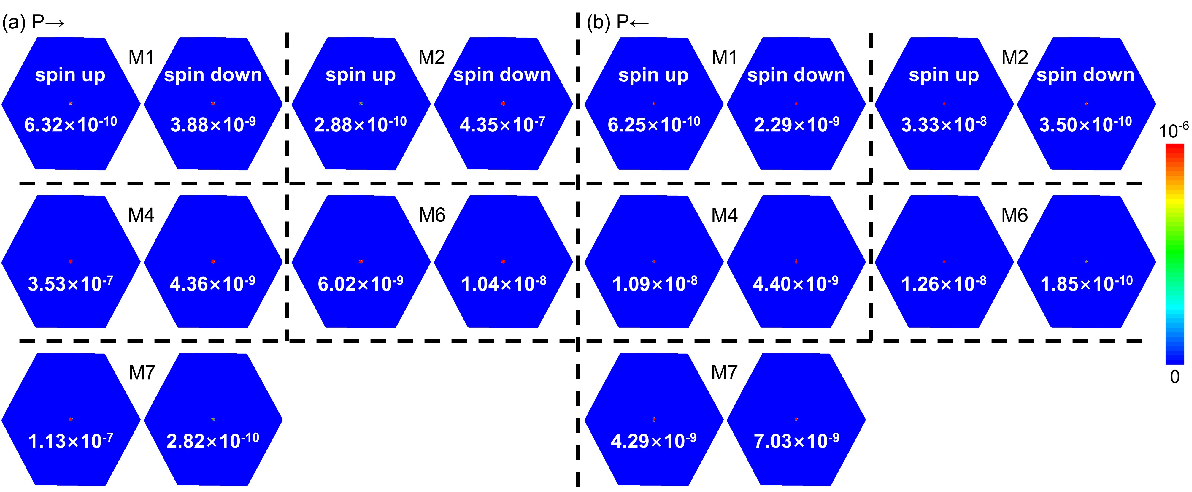}
	\caption{The spin-dependent $k_\Arrowvert$-resolved transmission spectra of the MBT-2L/IS/MBT-2L AMFTJs with (d) P$\rightarrow$ and (e) P$\leftarrow$.}
	\label{Fig11}
\end{figure*}

\begin{figure*}[htbp]
	\centering
	\includegraphics[width=17cm,angle=0]{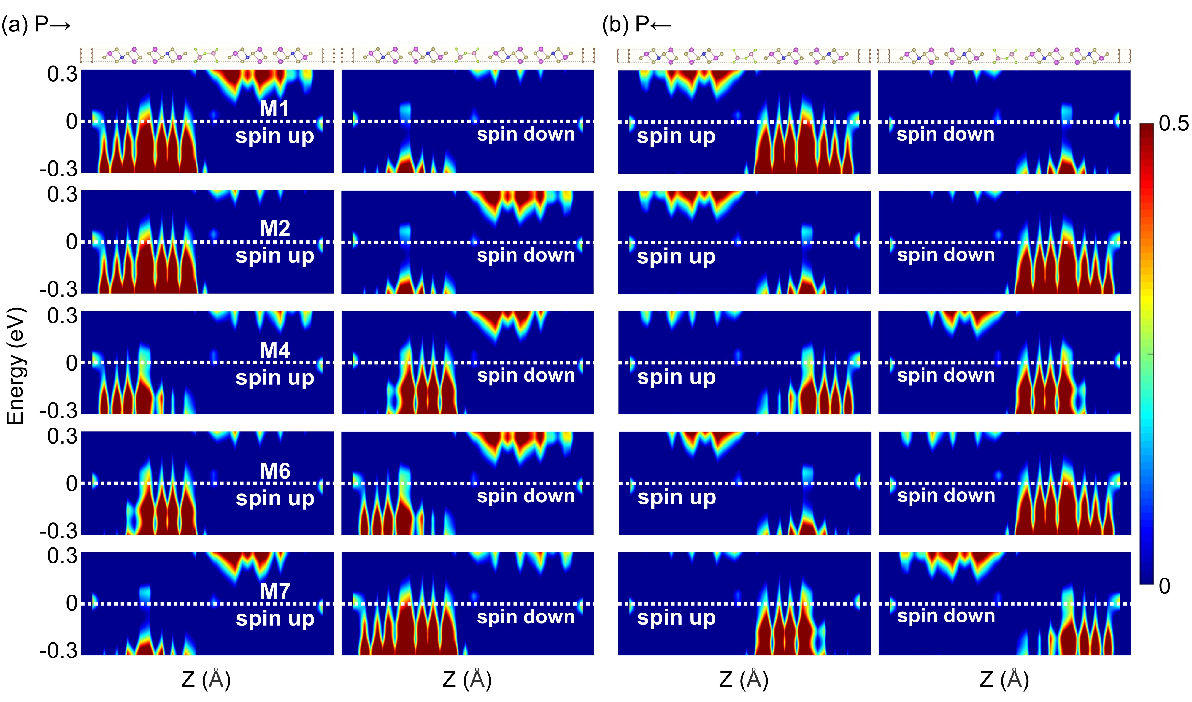}
	\caption{The averaged spin-resolved PDOS distribution over the $xy$ plane of the central scattering regions along the transport direction ($z$ axis) for MBT-2L/IS/MBT-2L AMFTJs in the equilibrium state. The white horizontal dashed line represents the Fermi level. (a) P $\rightarrow$; (b) P $\leftarrow$.}
	\label{Fig12}
\end{figure*}

\begin{figure*}[htbp]
	\centering
	\includegraphics[width=17cm,angle=0]{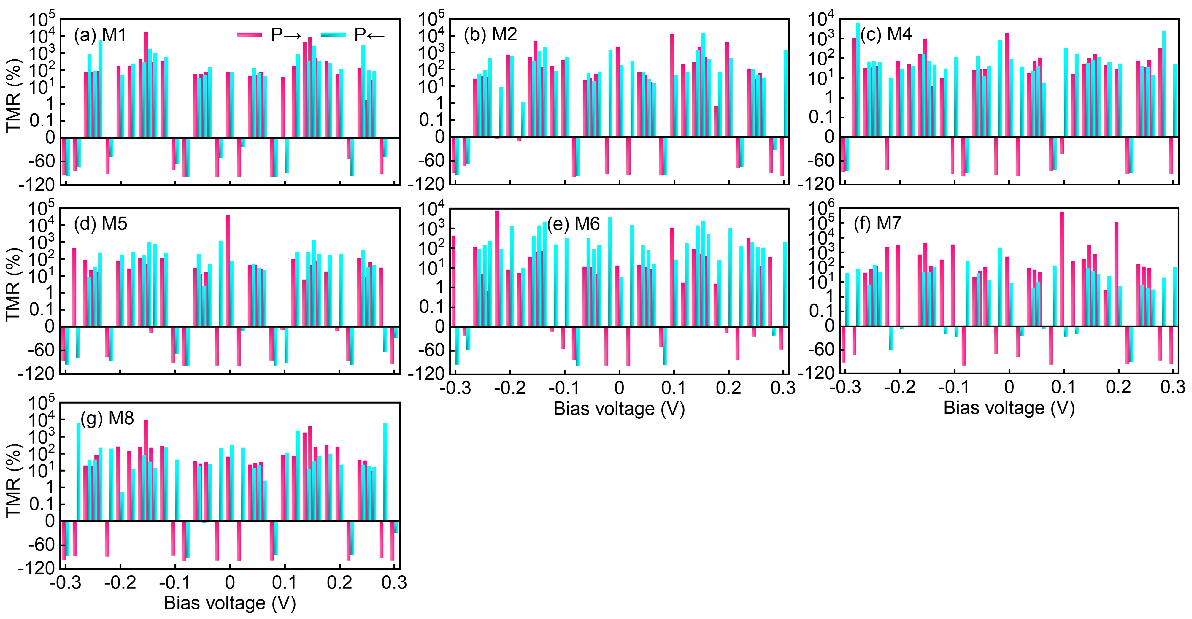}
	\caption{The variation of the TMR ratios (a)-(g) $vs$ the bias voltages for MBT-2L/IS/MBT-2L AMFTJs in different ferroelectric polarizations.}
	\label{Fig13}
\end{figure*}

\section{Summary}
In conclusion, our study demonstrates that the vdW AMFTJ, based on the MBT-2L/IS/MBT-2L structure, exhibits significant potential for next-generation non-volatile memory applications. The device shows remarkable spin-resolved electronic transport properties, achieving 16 distinct resistance states through manipulation of magnetization and ferroelectric polarization directions. In equilibrium, the tunneling magnetoresistance and electroresistance reach values of up to $3.79\times10^{4}$\% and $2.41\times10^{5}$\%, respectively, which can be further enhanced to $5.01\times10^{5}$\% and $4.97\times10^{5}$\% under external bias voltage. Additionally, the AMFTJ displays an excellent spin-filtering effect. These findings underscore the immense potential of the MBT-2L/IS/MBT-2L vdW AMFTJ for advancing antiferromagnetic spintronic devices.

\section{Acknowledgements}
This work was supported by the National Key Research and Development Program of China (Grant No. 2022YFB3505301), the National Natural Science Foundation of China (Grant No. 12304148), the Natural Science Basic Research Program of Shanxi (Grants No. 202203021222219, No. 202203021212393 and No. 20210302124252), and the Project funded by China Postdoctoral Science Foundation (Grant No.2023M731452).
\appendix

\section*{APPENDIX A: Band structure and electrostatic potential of MBT-2L/IS/MBT-2L}\label{sec:appendix_a}
The band structures of MBT-2L/IS/MBT-2L are shown in Fig.\ref{Fig7}. 
One can observe that the electronic band structure of the MBT-2L/IS/MBT-2L heterostructure in the M1 magnetic state shows almost no difference between the two ferroelectric polarization states [see Fig.\ref{Fig7}(a)]. This is because the magnetic alignment of MnBi$_2$Te$_4$ on both sides of the In$_2$Se$_3$ barrier layer is consistent. For the M2 configuration, the reversal of the ferroelectric polarization direction leads to a switch in the spin states in its electronic band structure [see Fig.\ref{Fig7}(b)].
The band structures of the other magnetic configurations exhibit similar behavior. The polarization field generated by the ferroelectric material causes spin splitting in the electronic band structure of the MBT-2L/IS/MBT-2L heterostructure (even when the net magnetic moment is zero). Reversing the ferroelectric polarization direction results in the exchange of spin orientation.

The electrostatic potential distribution of the MBT-2L/IS/MBT-2L heterostructure along the $z$ axis is given in Fig.\ref{Fig8}. It is clear that after the reversal of the ferroelectric polarization direction, the peak value of the electrostatic potential in the In$_2$Se$_3$ barrier layer is exchanged, demonstrating that the ferroelectricity of In$_2$Se$_3$ is maintained when forming a heterostructure with bilayer MnBi$_2$Te$_4$. This lays the prerequisite for the design of AMFTJ devices.

\section*{APPENDIX B: Graphene/MBT interface}\label{sec:appendix_B}
To study the interface stacking of MnBi$_2$Te$_4$ with graphene, we consider two different high-symmetry stacking configurations, as shown in Fig.\ref{Fig9}(a). One configuration has Te atoms located on top of C atoms, named Te-C, and the other has Te atoms located at the center of the hexagonal lattice of graphene, named Te-hollow. The total energy calculations demonstrate that Te-hollow is the lowest energy stacking configuration [see Fig.\ref{Fig9}(b)].

\section*{APPENDIX C: Electron transmission spectrum and PDOS in equilibrium state}\label{sec:appendix_c}
Spin-dependent transmission coefficient curves (Fig.\ref{Fig10}) and $k_\Arrowvert$-resolved transmission spectra (Fig.\ref{Fig11}) of the remaining magnetic configurations in the equilibrium state.
The equilibrium spin-resolved PDOS distribution is depicted in Fig.\ref{Fig12}
The results of this section have been discussed in detail in the main text and will not be repeated here.

\section*{APPENDIX D: TMR in non-equilibrium state}\label{sec:appendix_d}
Fig.\ref{Fig13} illustrates the TMR of distinct magnetic configurations within the tunnel junction under the influence of a bias voltage. A comparison of the TMR in the equilibrium state reveals that the bias voltage effectively modulates the TMR effect.

\bibliographystyle{apsrev4-2}
\bibliography{main}

\begin{thebibliography}{50}%
\makeatletter
\providecommand \@ifxundefined [1]{%
 \@ifx{#1\undefined}
}%
\providecommand \@ifnum [1]{%
 \ifnum #1\expandafter \@firstoftwo
 \else \expandafter \@secondoftwo
 \fi
}%
\providecommand \@ifx [1]{%
 \ifx #1\expandafter \@firstoftwo
 \else \expandafter \@secondoftwo
 \fi
}%
\providecommand \natexlab [1]{#1}%
\providecommand \enquote  [1]{``#1''}%
\providecommand \bibnamefont  [1]{#1}%
\providecommand \bibfnamefont [1]{#1}%
\providecommand \citenamefont [1]{#1}%
\providecommand \href@noop [0]{\@secondoftwo}%
\providecommand \href [0]{\begingroup \@sanitize@url \@href}%
\providecommand \@href[1]{\@@startlink{#1}\@@href}%
\providecommand \@@href[1]{\endgroup#1\@@endlink}%
\providecommand \@sanitize@url [0]{\catcode `\\12\catcode `\$12\catcode `\&12\catcode `\#12\catcode `\^12\catcode `\_12\catcode `\%12\relax}%
\providecommand \@@startlink[1]{}%
\providecommand \@@endlink[0]{}%
\providecommand \url  [0]{\begingroup\@sanitize@url \@url }%
\providecommand \@url [1]{\endgroup\@href {#1}{\urlprefix }}%
\providecommand \urlprefix  [0]{URL }%
\providecommand \Eprint [0]{\href }%
\providecommand \doibase [0]{https://doi.org/}%
\providecommand \selectlanguage [0]{\@gobble}%
\providecommand \bibinfo  [0]{\@secondoftwo}%
\providecommand \bibfield  [0]{\@secondoftwo}%
\providecommand \translation [1]{[#1]}%
\providecommand \BibitemOpen [0]{}%
\providecommand \bibitemStop [0]{}%
\providecommand \bibitemNoStop [0]{.\EOS\space}%
\providecommand \EOS [0]{\spacefactor3000\relax}%
\providecommand \BibitemShut  [1]{\csname bibitem#1\endcsname}%
\let\auto@bib@innerbib\@empty
\bibitem [{\citenamefont {Wolf}\ \emph {et~al.}(2001)\citenamefont {Wolf}, \citenamefont {Awschalom}, \citenamefont {Buhrman}, \citenamefont {Daughton}, \citenamefont {von Moln{\'a}r}, \citenamefont {Roukes}, \citenamefont {Chtchelkanova},\ and\ \citenamefont {Treger}}]{wolf2001spintronics}%
  \BibitemOpen
  \bibfield  {author} {\bibinfo {author} {\bibfnamefont {S.}~\bibnamefont {Wolf}}, \bibinfo {author} {\bibfnamefont {D.}~\bibnamefont {Awschalom}}, \bibinfo {author} {\bibfnamefont {R.}~\bibnamefont {Buhrman}}, \bibinfo {author} {\bibfnamefont {J.}~\bibnamefont {Daughton}}, \bibinfo {author} {\bibfnamefont {v.~S.}\ \bibnamefont {von Moln{\'a}r}}, \bibinfo {author} {\bibfnamefont {M.}~\bibnamefont {Roukes}}, \bibinfo {author} {\bibfnamefont {A.~Y.}\ \bibnamefont {Chtchelkanova}},\ and\ \bibinfo {author} {\bibfnamefont {D.}~\bibnamefont {Treger}},\ }\href {https://doi.org/10.1126/science.1065389} {\bibfield  {journal} {\bibinfo  {journal} {Science}\ }\textbf {\bibinfo {volume} {294}},\ \bibinfo {pages} {1488} (\bibinfo {year} {2001})}\BibitemShut {NoStop}%
\bibitem [{\citenamefont {{\v{Z}}uti{\'c}}\ \emph {et~al.}(2004)\citenamefont {{\v{Z}}uti{\'c}}, \citenamefont {Fabian},\ and\ \citenamefont {Sarma}}]{vzutic2004spintronics}%
  \BibitemOpen
  \bibfield  {author} {\bibinfo {author} {\bibfnamefont {I.}~\bibnamefont {{\v{Z}}uti{\'c}}}, \bibinfo {author} {\bibfnamefont {J.}~\bibnamefont {Fabian}},\ and\ \bibinfo {author} {\bibfnamefont {S.~D.}\ \bibnamefont {Sarma}},\ }\href {https://doi.org/10.1103/RevModPhys.76.323} {\bibfield  {journal} {\bibinfo  {journal} {Rev. Mod. Phys.}\ }\textbf {\bibinfo {volume} {76}},\ \bibinfo {pages} {323} (\bibinfo {year} {2004})}\BibitemShut {NoStop}%
\bibitem [{\citenamefont {Zhu}\ and\ \citenamefont {Park}(2006)}]{zhu2006magnetic}%
  \BibitemOpen
  \bibfield  {author} {\bibinfo {author} {\bibfnamefont {J.-G.~J.}\ \bibnamefont {Zhu}}\ and\ \bibinfo {author} {\bibfnamefont {C.}~\bibnamefont {Park}},\ }\href {https://doi.org/https://doi.org/10.1016/S1369-7021(06)71693-5} {\bibfield  {journal} {\bibinfo  {journal} {Mater. Today}\ }\textbf {\bibinfo {volume} {9}},\ \bibinfo {pages} {36} (\bibinfo {year} {2006})}\BibitemShut {NoStop}%
\bibitem [{\citenamefont {Ikeda}\ \emph {et~al.}(2007)\citenamefont {Ikeda}, \citenamefont {Hayakawa}, \citenamefont {Lee}, \citenamefont {Matsukura}, \citenamefont {Ohno}, \citenamefont {Hanyu},\ and\ \citenamefont {Ohno}}]{ikeda2007magnetic}%
  \BibitemOpen
  \bibfield  {author} {\bibinfo {author} {\bibfnamefont {S.}~\bibnamefont {Ikeda}}, \bibinfo {author} {\bibfnamefont {J.}~\bibnamefont {Hayakawa}}, \bibinfo {author} {\bibfnamefont {Y.~M.}\ \bibnamefont {Lee}}, \bibinfo {author} {\bibfnamefont {F.}~\bibnamefont {Matsukura}}, \bibinfo {author} {\bibfnamefont {Y.}~\bibnamefont {Ohno}}, \bibinfo {author} {\bibfnamefont {T.}~\bibnamefont {Hanyu}},\ and\ \bibinfo {author} {\bibfnamefont {H.}~\bibnamefont {Ohno}},\ }\href {https://doi.org/10.1109/TED.2007.894617} {\bibfield  {journal} {\bibinfo  {journal} {IEEE Trans. Electron Dev.}\ }\textbf {\bibinfo {volume} {54}},\ \bibinfo {pages} {991} (\bibinfo {year} {2007})}\BibitemShut {NoStop}%
\bibitem [{\citenamefont {Ralph}\ and\ \citenamefont {Stiles}(2008)}]{ralph2008spin}%
  \BibitemOpen
  \bibfield  {author} {\bibinfo {author} {\bibfnamefont {D.~C.}\ \bibnamefont {Ralph}}\ and\ \bibinfo {author} {\bibfnamefont {M.~D.}\ \bibnamefont {Stiles}},\ }\href {https://doi.org/https://doi.org/10.1016/j.jmmm.2007.12.019} {\bibfield  {journal} {\bibinfo  {journal} {J. Magn. Magn. Mater.}\ }\textbf {\bibinfo {volume} {320}},\ \bibinfo {pages} {1190} (\bibinfo {year} {2008})}\BibitemShut {NoStop}%
\bibitem [{\citenamefont {Kawahara}\ \emph {et~al.}(2012)\citenamefont {Kawahara}, \citenamefont {Ito}, \citenamefont {Takemura},\ and\ \citenamefont {Ohno}}]{kawahara2012spin}%
  \BibitemOpen
  \bibfield  {author} {\bibinfo {author} {\bibfnamefont {T.}~\bibnamefont {Kawahara}}, \bibinfo {author} {\bibfnamefont {K.}~\bibnamefont {Ito}}, \bibinfo {author} {\bibfnamefont {R.}~\bibnamefont {Takemura}},\ and\ \bibinfo {author} {\bibfnamefont {H.}~\bibnamefont {Ohno}},\ }\href {https://doi.org/https://doi.org/10.1016/j.microrel.2011.09.028} {\bibfield  {journal} {\bibinfo  {journal} {Microelectron. Reliab.}\ }\textbf {\bibinfo {volume} {52}},\ \bibinfo {pages} {613} (\bibinfo {year} {2012})}\BibitemShut {NoStop}%
\bibitem [{\citenamefont {Han}\ \emph {et~al.}(2023)\citenamefont {Han}, \citenamefont {Cheng}, \citenamefont {Liu}, \citenamefont {Ohno},\ and\ \citenamefont {Fukami}}]{han2023coherent}%
  \BibitemOpen
  \bibfield  {author} {\bibinfo {author} {\bibfnamefont {J.}~\bibnamefont {Han}}, \bibinfo {author} {\bibfnamefont {R.}~\bibnamefont {Cheng}}, \bibinfo {author} {\bibfnamefont {L.}~\bibnamefont {Liu}}, \bibinfo {author} {\bibfnamefont {H.}~\bibnamefont {Ohno}},\ and\ \bibinfo {author} {\bibfnamefont {S.}~\bibnamefont {Fukami}},\ }\href {https://doi.org/https://doi.org/10.1038/s41563-023-01492-6} {\bibfield  {journal} {\bibinfo  {journal} {Nat. Mater.}\ }\textbf {\bibinfo {volume} {22}},\ \bibinfo {pages} {684} (\bibinfo {year} {2023})}\BibitemShut {NoStop}%
\bibitem [{\citenamefont {Dal~Din}\ \emph {et~al.}(2024)\citenamefont {Dal~Din}, \citenamefont {Amin}, \citenamefont {Wadley},\ and\ \citenamefont {Edmonds}}]{dal2024antiferromagnetic}%
  \BibitemOpen
  \bibfield  {author} {\bibinfo {author} {\bibfnamefont {A.}~\bibnamefont {Dal~Din}}, \bibinfo {author} {\bibfnamefont {O.}~\bibnamefont {Amin}}, \bibinfo {author} {\bibfnamefont {P.}~\bibnamefont {Wadley}},\ and\ \bibinfo {author} {\bibfnamefont {K.}~\bibnamefont {Edmonds}},\ }\href {https://doi.org/https://doi.org/10.1038/s44306-024-00029-0} {\bibfield  {journal} {\bibinfo  {journal} {Npj Spintron.}\ }\textbf {\bibinfo {volume} {2}},\ \bibinfo {pages} {25} (\bibinfo {year} {2024})}\BibitemShut {NoStop}%
\bibitem [{\citenamefont {Qin}\ \emph {et~al.}(2023)\citenamefont {Qin}, \citenamefont {Yan}, \citenamefont {Wang}, \citenamefont {Chen}, \citenamefont {Meng}, \citenamefont {Dong}, \citenamefont {Zhu}, \citenamefont {Cai}, \citenamefont {Feng}, \citenamefont {Zhou} \emph {et~al.}}]{qin2023room}%
  \BibitemOpen
  \bibfield  {author} {\bibinfo {author} {\bibfnamefont {P.}~\bibnamefont {Qin}}, \bibinfo {author} {\bibfnamefont {H.}~\bibnamefont {Yan}}, \bibinfo {author} {\bibfnamefont {X.}~\bibnamefont {Wang}}, \bibinfo {author} {\bibfnamefont {H.}~\bibnamefont {Chen}}, \bibinfo {author} {\bibfnamefont {Z.}~\bibnamefont {Meng}}, \bibinfo {author} {\bibfnamefont {J.}~\bibnamefont {Dong}}, \bibinfo {author} {\bibfnamefont {M.}~\bibnamefont {Zhu}}, \bibinfo {author} {\bibfnamefont {J.}~\bibnamefont {Cai}}, \bibinfo {author} {\bibfnamefont {Z.}~\bibnamefont {Feng}}, \bibinfo {author} {\bibfnamefont {X.}~\bibnamefont {Zhou}}, \emph {et~al.},\ }\href {https://doi.org/https://doi.org/10.1038/s41586-022-05461-y} {\bibfield  {journal} {\bibinfo  {journal} {Nature}\ }\textbf {\bibinfo {volume} {613}},\ \bibinfo {pages} {485} (\bibinfo {year} {2023})}\BibitemShut {NoStop}%
\bibitem [{\citenamefont {Gu}\ \emph {et~al.}(2023)\citenamefont {Gu}, \citenamefont {Wang}, \citenamefont {Su}, \citenamefont {Dong}, \citenamefont {Wang}, \citenamefont {Han}, \citenamefont {Watanabe}, \citenamefont {Taniguchi}, \citenamefont {Ji}, \citenamefont {Sun} \emph {et~al.}}]{gu2023multi}%
  \BibitemOpen
  \bibfield  {author} {\bibinfo {author} {\bibfnamefont {P.}~\bibnamefont {Gu}}, \bibinfo {author} {\bibfnamefont {C.}~\bibnamefont {Wang}}, \bibinfo {author} {\bibfnamefont {D.}~\bibnamefont {Su}}, \bibinfo {author} {\bibfnamefont {Z.}~\bibnamefont {Dong}}, \bibinfo {author} {\bibfnamefont {Q.}~\bibnamefont {Wang}}, \bibinfo {author} {\bibfnamefont {Z.}~\bibnamefont {Han}}, \bibinfo {author} {\bibfnamefont {K.}~\bibnamefont {Watanabe}}, \bibinfo {author} {\bibfnamefont {T.}~\bibnamefont {Taniguchi}}, \bibinfo {author} {\bibfnamefont {W.}~\bibnamefont {Ji}}, \bibinfo {author} {\bibfnamefont {Y.}~\bibnamefont {Sun}}, \emph {et~al.},\ }\href {https://doi.org/https://doi.org/10.1038/s41467-023-39004-4} {\bibfield  {journal} {\bibinfo  {journal} {Nat. Commun.}\ }\textbf {\bibinfo {volume} {14}},\ \bibinfo {pages} {3221} (\bibinfo {year} {2023})}\BibitemShut {NoStop}%
\bibitem [{\citenamefont {Zhai}\ \emph {et~al.}(2021)\citenamefont {Zhai}, \citenamefont {Xu}, \citenamefont {Cui}, \citenamefont {Zhu}, \citenamefont {Yang},\ and\ \citenamefont {Blanter}}]{zhai2021electrically}%
  \BibitemOpen
  \bibfield  {author} {\bibinfo {author} {\bibfnamefont {X.}~\bibnamefont {Zhai}}, \bibinfo {author} {\bibfnamefont {Z.}~\bibnamefont {Xu}}, \bibinfo {author} {\bibfnamefont {Q.}~\bibnamefont {Cui}}, \bibinfo {author} {\bibfnamefont {Y.}~\bibnamefont {Zhu}}, \bibinfo {author} {\bibfnamefont {H.}~\bibnamefont {Yang}},\ and\ \bibinfo {author} {\bibfnamefont {Y.~M.}\ \bibnamefont {Blanter}},\ }\href {https://doi.org/10.1103/PhysRevApplied.16.014032} {\bibfield  {journal} {\bibinfo  {journal} {Phys. Rev. Appl.}\ }\textbf {\bibinfo {volume} {16}},\ \bibinfo {pages} {014032} (\bibinfo {year} {2021})}\BibitemShut {NoStop}%
\bibitem [{\citenamefont {Chirac}\ \emph {et~al.}(2020)\citenamefont {Chirac}, \citenamefont {Chauleau}, \citenamefont {Thibaudeau}, \citenamefont {Gomonay},\ and\ \citenamefont {Viret}}]{chirac2020ultrafast}%
  \BibitemOpen
  \bibfield  {author} {\bibinfo {author} {\bibfnamefont {T.}~\bibnamefont {Chirac}}, \bibinfo {author} {\bibfnamefont {J.-Y.}\ \bibnamefont {Chauleau}}, \bibinfo {author} {\bibfnamefont {P.}~\bibnamefont {Thibaudeau}}, \bibinfo {author} {\bibfnamefont {O.}~\bibnamefont {Gomonay}},\ and\ \bibinfo {author} {\bibfnamefont {M.}~\bibnamefont {Viret}},\ }\href {https://doi.org/10.1103/PhysRevB.102.134415} {\bibfield  {journal} {\bibinfo  {journal} {Phys. Rev. B}\ }\textbf {\bibinfo {volume} {102}},\ \bibinfo {pages} {134415} (\bibinfo {year} {2020})}\BibitemShut {NoStop}%
\bibitem [{\citenamefont {Wang}\ \emph {et~al.}(2023)\citenamefont {Wang}, \citenamefont {Zhou}, \citenamefont {Xu}, \citenamefont {Zhang}, \citenamefont {Zhu}, \citenamefont {Guo}, \citenamefont {Deng}, \citenamefont {Yang}, \citenamefont {Meng}, \citenamefont {He} \emph {et~al.}}]{wang2023field}%
  \BibitemOpen
  \bibfield  {author} {\bibinfo {author} {\bibfnamefont {M.}~\bibnamefont {Wang}}, \bibinfo {author} {\bibfnamefont {J.}~\bibnamefont {Zhou}}, \bibinfo {author} {\bibfnamefont {X.}~\bibnamefont {Xu}}, \bibinfo {author} {\bibfnamefont {T.}~\bibnamefont {Zhang}}, \bibinfo {author} {\bibfnamefont {Z.}~\bibnamefont {Zhu}}, \bibinfo {author} {\bibfnamefont {Z.}~\bibnamefont {Guo}}, \bibinfo {author} {\bibfnamefont {Y.}~\bibnamefont {Deng}}, \bibinfo {author} {\bibfnamefont {M.}~\bibnamefont {Yang}}, \bibinfo {author} {\bibfnamefont {K.}~\bibnamefont {Meng}}, \bibinfo {author} {\bibfnamefont {B.}~\bibnamefont {He}}, \emph {et~al.},\ }\href {https://doi.org/https://doi.org/10.1038/s41467-023-38550-1} {\bibfield  {journal} {\bibinfo  {journal} {Nat. Commun.}\ }\textbf {\bibinfo {volume} {14}},\ \bibinfo {pages} {2871} (\bibinfo {year} {2023})}\BibitemShut {NoStop}%
\bibitem [{\citenamefont {Chen}\ \emph {et~al.}(2023)\citenamefont {Chen}, \citenamefont {Higo}, \citenamefont {Tanaka}, \citenamefont {Nomoto}, \citenamefont {Tsai}, \citenamefont {Idzuchi}, \citenamefont {Shiga}, \citenamefont {Sakamoto}, \citenamefont {Ando}, \citenamefont {Kosaki} \emph {et~al.}}]{chen2023octupole}%
  \BibitemOpen
  \bibfield  {author} {\bibinfo {author} {\bibfnamefont {X.}~\bibnamefont {Chen}}, \bibinfo {author} {\bibfnamefont {T.}~\bibnamefont {Higo}}, \bibinfo {author} {\bibfnamefont {K.}~\bibnamefont {Tanaka}}, \bibinfo {author} {\bibfnamefont {T.}~\bibnamefont {Nomoto}}, \bibinfo {author} {\bibfnamefont {H.}~\bibnamefont {Tsai}}, \bibinfo {author} {\bibfnamefont {H.}~\bibnamefont {Idzuchi}}, \bibinfo {author} {\bibfnamefont {M.}~\bibnamefont {Shiga}}, \bibinfo {author} {\bibfnamefont {S.}~\bibnamefont {Sakamoto}}, \bibinfo {author} {\bibfnamefont {R.}~\bibnamefont {Ando}}, \bibinfo {author} {\bibfnamefont {H.}~\bibnamefont {Kosaki}}, \emph {et~al.},\ }\href {https://doi.org/https://doi.org/10.1038/s41586-022-05463-w} {\bibfield  {journal} {\bibinfo  {journal} {Nature}\ }\textbf {\bibinfo {volume} {613}},\ \bibinfo {pages} {490} (\bibinfo {year} {2023})}\BibitemShut {NoStop}%
\bibitem [{\citenamefont {Zhang}\ \emph {et~al.}(2024)\citenamefont {Zhang}, \citenamefont {Lang}, \citenamefont {Zhu}, \citenamefont {Li}, \citenamefont {Zhao}, \citenamefont {Wei},\ and\ \citenamefont {Zhou}}]{zhang2024electronic}%
  \BibitemOpen
  \bibfield  {author} {\bibinfo {author} {\bibfnamefont {Z.-W.}\ \bibnamefont {Zhang}}, \bibinfo {author} {\bibfnamefont {Y.-F.}\ \bibnamefont {Lang}}, \bibinfo {author} {\bibfnamefont {H.-P.}\ \bibnamefont {Zhu}}, \bibinfo {author} {\bibfnamefont {B.}~\bibnamefont {Li}}, \bibinfo {author} {\bibfnamefont {Y.-Q.}\ \bibnamefont {Zhao}}, \bibinfo {author} {\bibfnamefont {B.}~\bibnamefont {Wei}},\ and\ \bibinfo {author} {\bibfnamefont {W.-X.}\ \bibnamefont {Zhou}},\ }\href {https://doi.org/10.1103/PhysRevApplied.21.064012} {\bibfield  {journal} {\bibinfo  {journal} {Phys. Rev. Appl.}\ }\textbf {\bibinfo {volume} {21}},\ \bibinfo {pages} {064012} (\bibinfo {year} {2024})}\BibitemShut {NoStop}%
\bibitem [{\citenamefont {Dong}\ \emph {et~al.}(2023)\citenamefont {Dong}, \citenamefont {Jia}, \citenamefont {Yan}, \citenamefont {Shen}, \citenamefont {Li}, \citenamefont {Qiao},\ and\ \citenamefont {Xu}}]{0dong2023spin}%
  \BibitemOpen
  \bibfield  {author} {\bibinfo {author} {\bibfnamefont {X.}~\bibnamefont {Dong}}, \bibinfo {author} {\bibfnamefont {X.}~\bibnamefont {Jia}}, \bibinfo {author} {\bibfnamefont {Z.}~\bibnamefont {Yan}}, \bibinfo {author} {\bibfnamefont {X.}~\bibnamefont {Shen}}, \bibinfo {author} {\bibfnamefont {Z.}~\bibnamefont {Li}}, \bibinfo {author} {\bibfnamefont {Z.}~\bibnamefont {Qiao}},\ and\ \bibinfo {author} {\bibfnamefont {X.}~\bibnamefont {Xu}},\ }\href {https://doi.org/10.1088/0256-307x/40/8/087301} {\bibfield  {journal} {\bibinfo  {journal} {Chin. Phys. Lett.}\ }\textbf {\bibinfo {volume} {40}},\ \bibinfo {pages} {087301} (\bibinfo {year} {2023})}\BibitemShut {NoStop}%
\bibitem [{\citenamefont {Su}\ \emph {et~al.}(2020)\citenamefont {Su}, \citenamefont {Li}, \citenamefont {Zhu}, \citenamefont {Zhang}, \citenamefont {You},\ and\ \citenamefont {Tsymbal}}]{0su2020van}%
  \BibitemOpen
  \bibfield  {author} {\bibinfo {author} {\bibfnamefont {Y.}~\bibnamefont {Su}}, \bibinfo {author} {\bibfnamefont {X.}~\bibnamefont {Li}}, \bibinfo {author} {\bibfnamefont {M.}~\bibnamefont {Zhu}}, \bibinfo {author} {\bibfnamefont {J.}~\bibnamefont {Zhang}}, \bibinfo {author} {\bibfnamefont {L.}~\bibnamefont {You}},\ and\ \bibinfo {author} {\bibfnamefont {E.~Y.}\ \bibnamefont {Tsymbal}},\ }\href {https://doi.org/https://doi.org/10.1021/acs.nanolett.0c03452} {\bibfield  {journal} {\bibinfo  {journal} {Nano Lett.}\ }\textbf {\bibinfo {volume} {21}},\ \bibinfo {pages} {175} (\bibinfo {year} {2020})}\BibitemShut {NoStop}%
\bibitem [{\citenamefont {Feng}\ \emph {et~al.}(2024)\citenamefont {Feng}, \citenamefont {Han}, \citenamefont {Zhang}, \citenamefont {Lin}, \citenamefont {Gao}, \citenamefont {Yang},\ and\ \citenamefont {Meng}}]{0feng2024van}%
  \BibitemOpen
  \bibfield  {author} {\bibinfo {author} {\bibfnamefont {Y.}~\bibnamefont {Feng}}, \bibinfo {author} {\bibfnamefont {J.}~\bibnamefont {Han}}, \bibinfo {author} {\bibfnamefont {K.}~\bibnamefont {Zhang}}, \bibinfo {author} {\bibfnamefont {X.}~\bibnamefont {Lin}}, \bibinfo {author} {\bibfnamefont {G.}~\bibnamefont {Gao}}, \bibinfo {author} {\bibfnamefont {Q.}~\bibnamefont {Yang}},\ and\ \bibinfo {author} {\bibfnamefont {S.}~\bibnamefont {Meng}},\ }\href {https://doi.org/10.1103/PhysRevB.109.085433} {\bibfield  {journal} {\bibinfo  {journal} {Phys. Rev. B}\ }\textbf {\bibinfo {volume} {109}},\ \bibinfo {pages} {085433} (\bibinfo {year} {2024})}\BibitemShut {NoStop}%
\bibitem [{\citenamefont {Yan}\ \emph {et~al.}(2020)\citenamefont {Yan}, \citenamefont {Zhang}, \citenamefont {Dong}, \citenamefont {Qi},\ and\ \citenamefont {Xu}}]{0yan2020significant}%
  \BibitemOpen
  \bibfield  {author} {\bibinfo {author} {\bibfnamefont {Z.}~\bibnamefont {Yan}}, \bibinfo {author} {\bibfnamefont {R.}~\bibnamefont {Zhang}}, \bibinfo {author} {\bibfnamefont {X.}~\bibnamefont {Dong}}, \bibinfo {author} {\bibfnamefont {S.}~\bibnamefont {Qi}},\ and\ \bibinfo {author} {\bibfnamefont {X.}~\bibnamefont {Xu}},\ }\href {https://doi.org/10.1039/D0CP02534H} {\bibfield  {journal} {\bibinfo  {journal} {Phys. Chem. Chem. Phys.}\ }\textbf {\bibinfo {volume} {22}},\ \bibinfo {pages} {14773} (\bibinfo {year} {2020})}\BibitemShut {NoStop}%
\bibitem [{\citenamefont {Yang}\ \emph {et~al.}(2024)\citenamefont {Yang}, \citenamefont {Wu}, \citenamefont {Zhao}, \citenamefont {Liu}, \citenamefont {Lu}, \citenamefont {Li},\ and\ \citenamefont {Yang}}]{0yang2024multistate}%
  \BibitemOpen
  \bibfield  {author} {\bibinfo {author} {\bibfnamefont {J.}~\bibnamefont {Yang}}, \bibinfo {author} {\bibfnamefont {B.}~\bibnamefont {Wu}}, \bibinfo {author} {\bibfnamefont {S.}~\bibnamefont {Zhao}}, \bibinfo {author} {\bibfnamefont {S.}~\bibnamefont {Liu}}, \bibinfo {author} {\bibfnamefont {J.}~\bibnamefont {Lu}}, \bibinfo {author} {\bibfnamefont {S.}~\bibnamefont {Li}},\ and\ \bibinfo {author} {\bibfnamefont {J.}~\bibnamefont {Yang}},\ }\href {https://doi.org/10.1103/PhysRevApplied.22.014017} {\bibfield  {journal} {\bibinfo  {journal} {Phys. Rev. Appl.}\ }\textbf {\bibinfo {volume} {22}},\ \bibinfo {pages} {014017} (\bibinfo {year} {2024})}\BibitemShut {NoStop}%
\bibitem [{\citenamefont {Yu}\ \emph {et~al.}(2023)\citenamefont {Yu}, \citenamefont {Zhang},\ and\ \citenamefont {Wang}}]{0yu2023fully}%
  \BibitemOpen
  \bibfield  {author} {\bibinfo {author} {\bibfnamefont {X.}~\bibnamefont {Yu}}, \bibinfo {author} {\bibfnamefont {X.}~\bibnamefont {Zhang}},\ and\ \bibinfo {author} {\bibfnamefont {J.}~\bibnamefont {Wang}},\ }\href {https://doi.org/10.1021/acsnano.3c08747} {\bibfield  {journal} {\bibinfo  {journal} {ACS nano}\ }\textbf {\bibinfo {volume} {17}},\ \bibinfo {pages} {25348} (\bibinfo {year} {2023})}\BibitemShut {NoStop}%
\bibitem [{\citenamefont {Yan}\ \emph {et~al.}(2021)\citenamefont {Yan}, \citenamefont {Jia}, \citenamefont {Shi}, \citenamefont {Dong},\ and\ \citenamefont {Xu}}]{0yan2021barrier}%
  \BibitemOpen
  \bibfield  {author} {\bibinfo {author} {\bibfnamefont {Z.}~\bibnamefont {Yan}}, \bibinfo {author} {\bibfnamefont {X.}~\bibnamefont {Jia}}, \bibinfo {author} {\bibfnamefont {X.}~\bibnamefont {Shi}}, \bibinfo {author} {\bibfnamefont {X.}~\bibnamefont {Dong}},\ and\ \bibinfo {author} {\bibfnamefont {X.}~\bibnamefont {Xu}},\ }\href {https://doi.org/10.1063/5.0052720} {\bibfield  {journal} {\bibinfo  {journal} {Appl. Phys. Lett.}\ }\textbf {\bibinfo {volume} {118}},\ \bibinfo {pages} {223503} (\bibinfo {year} {2021})}\BibitemShut {NoStop}%
\bibitem [{\citenamefont {Zhang}\ \emph {et~al.}(2021)\citenamefont {Zhang}, \citenamefont {Zhou}, \citenamefont {Li}, \citenamefont {Shen},\ and\ \citenamefont {Feng}}]{0zhang2021recent}%
  \BibitemOpen
  \bibfield  {author} {\bibinfo {author} {\bibfnamefont {L.}~\bibnamefont {Zhang}}, \bibinfo {author} {\bibfnamefont {J.}~\bibnamefont {Zhou}}, \bibinfo {author} {\bibfnamefont {H.}~\bibnamefont {Li}}, \bibinfo {author} {\bibfnamefont {L.}~\bibnamefont {Shen}},\ and\ \bibinfo {author} {\bibfnamefont {Y.~P.}\ \bibnamefont {Feng}},\ }\href {https://doi.org/https://doi.org/10.1063/5.0032538} {\bibfield  {journal} {\bibinfo  {journal} {Appl. Phys. Rev.}\ }\textbf {\bibinfo {volume} {8}},\ \bibinfo {pages} {021308} (\bibinfo {year} {2021})}\BibitemShut {NoStop}%
\bibitem [{\citenamefont {Zhu}\ \emph {et~al.}(2021)\citenamefont {Zhu}, \citenamefont {Guo}, \citenamefont {Jiang}, \citenamefont {Yan}, \citenamefont {Yan},\ and\ \citenamefont {Han}}]{0zhu2021giant}%
  \BibitemOpen
  \bibfield  {author} {\bibinfo {author} {\bibfnamefont {Y.}~\bibnamefont {Zhu}}, \bibinfo {author} {\bibfnamefont {X.}~\bibnamefont {Guo}}, \bibinfo {author} {\bibfnamefont {L.}~\bibnamefont {Jiang}}, \bibinfo {author} {\bibfnamefont {Z.}~\bibnamefont {Yan}}, \bibinfo {author} {\bibfnamefont {Y.}~\bibnamefont {Yan}},\ and\ \bibinfo {author} {\bibfnamefont {X.}~\bibnamefont {Han}},\ }\href {https://doi.org/10.1103/PhysRevB.103.134437} {\bibfield  {journal} {\bibinfo  {journal} {Phys. Rev. B}\ }\textbf {\bibinfo {volume} {103}},\ \bibinfo {pages} {134437} (\bibinfo {year} {2021})}\BibitemShut {NoStop}%
\bibitem [{\citenamefont {Yan}\ \emph {et~al.}(2024)\citenamefont {Yan}, \citenamefont {Yang}, \citenamefont {Fang}, \citenamefont {Lu},\ and\ \citenamefont {Xu}}]{0yan2024giant}%
  \BibitemOpen
  \bibfield  {author} {\bibinfo {author} {\bibfnamefont {Z.}~\bibnamefont {Yan}}, \bibinfo {author} {\bibfnamefont {R.}~\bibnamefont {Yang}}, \bibinfo {author} {\bibfnamefont {C.}~\bibnamefont {Fang}}, \bibinfo {author} {\bibfnamefont {W.}~\bibnamefont {Lu}},\ and\ \bibinfo {author} {\bibfnamefont {X.}~\bibnamefont {Xu}},\ }\href {https://doi.org/10.1103/physrevb.109.205409} {\bibfield  {journal} {\bibinfo  {journal} {Phys. Rev. B}\ }\textbf {\bibinfo {volume} {109}},\ \bibinfo {pages} {205409} (\bibinfo {year} {2024})}\BibitemShut {NoStop}%
\bibitem [{\citenamefont {Zhang}\ \emph {et~al.}(2023)\citenamefont {Zhang}, \citenamefont {Li}, \citenamefont {Jiang}, \citenamefont {Wang}, \citenamefont {Li},\ and\ \citenamefont {Ghosh}}]{0zhang2023current}%
  \BibitemOpen
  \bibfield  {author} {\bibinfo {author} {\bibfnamefont {L.}~\bibnamefont {Zhang}}, \bibinfo {author} {\bibfnamefont {H.}~\bibnamefont {Li}}, \bibinfo {author} {\bibfnamefont {Y.}~\bibnamefont {Jiang}}, \bibinfo {author} {\bibfnamefont {Z.}~\bibnamefont {Wang}}, \bibinfo {author} {\bibfnamefont {T.}~\bibnamefont {Li}},\ and\ \bibinfo {author} {\bibfnamefont {S.}~\bibnamefont {Ghosh}},\ }\href {https://doi.org/10.1103/PhysRevApplied.20.044056} {\bibfield  {journal} {\bibinfo  {journal} {Phys. Rev. Appl.}\ }\textbf {\bibinfo {volume} {20}},\ \bibinfo {pages} {044056} (\bibinfo {year} {2023})}\BibitemShut {NoStop}%
\bibitem [{\citenamefont {Yan}\ \emph {et~al.}(2022)\citenamefont {Yan}, \citenamefont {Li}, \citenamefont {Han}, \citenamefont {Qiao},\ and\ \citenamefont {Xu}}]{50yan2022giant}%
  \BibitemOpen
  \bibfield  {author} {\bibinfo {author} {\bibfnamefont {Z.}~\bibnamefont {Yan}}, \bibinfo {author} {\bibfnamefont {Z.}~\bibnamefont {Li}}, \bibinfo {author} {\bibfnamefont {Y.}~\bibnamefont {Han}}, \bibinfo {author} {\bibfnamefont {Z.}~\bibnamefont {Qiao}},\ and\ \bibinfo {author} {\bibfnamefont {X.}~\bibnamefont {Xu}},\ }\href {https://doi.org/10.1103/physrevb.105.075423} {\bibfield  {journal} {\bibinfo  {journal} {Phys. Rev. B}\ }\textbf {\bibinfo {volume} {105}},\ \bibinfo {pages} {075423} (\bibinfo {year} {2022})}\BibitemShut {NoStop}%
\bibitem [{\citenamefont {Kresse}\ and\ \citenamefont {Furthm{\"u}ller}(1996)}]{31kresse1996efficient}%
  \BibitemOpen
  \bibfield  {author} {\bibinfo {author} {\bibfnamefont {G.}~\bibnamefont {Kresse}}\ and\ \bibinfo {author} {\bibfnamefont {J.}~\bibnamefont {Furthm{\"u}ller}},\ }\href {https://doi.org/10.1103/physrevb.54.11169} {\bibfield  {journal} {\bibinfo  {journal} {Phys. Rev. B}\ }\textbf {\bibinfo {volume} {54}},\ \bibinfo {pages} {11169} (\bibinfo {year} {1996})}\BibitemShut {NoStop}%
\bibitem [{\citenamefont {Bl{\"o}chl}(1994)}]{32blochl1994projector}%
  \BibitemOpen
  \bibfield  {author} {\bibinfo {author} {\bibfnamefont {P.~E.}\ \bibnamefont {Bl{\"o}chl}},\ }\href {https://doi.org/10.1103/physrevb.50.17953} {\bibfield  {journal} {\bibinfo  {journal} {Phys. Rev. B}\ }\textbf {\bibinfo {volume} {50}},\ \bibinfo {pages} {17953} (\bibinfo {year} {1994})}\BibitemShut {NoStop}%
\bibitem [{\citenamefont {Perdew}\ \emph {et~al.}(1992)\citenamefont {Perdew}, \citenamefont {Chevary}, \citenamefont {Vosko}, \citenamefont {Jackson}, \citenamefont {Pederson}, \citenamefont {Singh},\ and\ \citenamefont {Fiolhais}}]{33perdew1992atoms}%
  \BibitemOpen
  \bibfield  {author} {\bibinfo {author} {\bibfnamefont {J.~P.}\ \bibnamefont {Perdew}}, \bibinfo {author} {\bibfnamefont {J.~A.}\ \bibnamefont {Chevary}}, \bibinfo {author} {\bibfnamefont {S.~H.}\ \bibnamefont {Vosko}}, \bibinfo {author} {\bibfnamefont {K.~A.}\ \bibnamefont {Jackson}}, \bibinfo {author} {\bibfnamefont {M.~R.}\ \bibnamefont {Pederson}}, \bibinfo {author} {\bibfnamefont {D.~J.}\ \bibnamefont {Singh}},\ and\ \bibinfo {author} {\bibfnamefont {C.}~\bibnamefont {Fiolhais}},\ }\href {https://doi.org/10.1103/physrevb.46.6671} {\bibfield  {journal} {\bibinfo  {journal} {Phys. Rev. B}\ }\textbf {\bibinfo {volume} {46}},\ \bibinfo {pages} {6671} (\bibinfo {year} {1992})}\BibitemShut {NoStop}%
\bibitem [{\citenamefont {Johnson}\ and\ \citenamefont {Becke}(2005)}]{34johnson2005post}%
  \BibitemOpen
  \bibfield  {author} {\bibinfo {author} {\bibfnamefont {E.~R.}\ \bibnamefont {Johnson}}\ and\ \bibinfo {author} {\bibfnamefont {A.~D.}\ \bibnamefont {Becke}},\ }\href {https://doi.org/10.1063/1.1949201} {\bibfield  {journal} {\bibinfo  {journal} {J. Chem. Phys.}\ }\textbf {\bibinfo {volume} {123}},\ \bibinfo {pages} {024101} (\bibinfo {year} {2005})}\BibitemShut {NoStop}%
\bibitem [{\citenamefont {Monkhorst}\ and\ \citenamefont {Pack}(1976)}]{35monkhorst1976special}%
  \BibitemOpen
  \bibfield  {author} {\bibinfo {author} {\bibfnamefont {H.~J.}\ \bibnamefont {Monkhorst}}\ and\ \bibinfo {author} {\bibfnamefont {J.~D.}\ \bibnamefont {Pack}},\ }\href {https://doi.org/10.1103/PhysRevB.13.5188} {\bibfield  {journal} {\bibinfo  {journal} {Phys. Rev. B}\ }\textbf {\bibinfo {volume} {13}},\ \bibinfo {pages} {5188} (\bibinfo {year} {1976})}\BibitemShut {NoStop}%
\bibitem [{\citenamefont {Liechtenstein}\ \emph {et~al.}(1995)\citenamefont {Liechtenstein}, \citenamefont {Anisimov},\ and\ \citenamefont {Zaanen}}]{liechtenstein1995density}%
  \BibitemOpen
  \bibfield  {author} {\bibinfo {author} {\bibfnamefont {A.}~\bibnamefont {Liechtenstein}}, \bibinfo {author} {\bibfnamefont {V.~I.}\ \bibnamefont {Anisimov}},\ and\ \bibinfo {author} {\bibfnamefont {J.}~\bibnamefont {Zaanen}},\ }\href {https://doi.org/10.1103/PhysRevB.52.R5467} {\bibfield  {journal} {\bibinfo  {journal} {Phys. Rev. B}\ }\textbf {\bibinfo {volume} {52}},\ \bibinfo {pages} {R5467} (\bibinfo {year} {1995})}\BibitemShut {NoStop}%
\bibitem [{\citenamefont {King-Smith}\ and\ \citenamefont {Vanderbilt}(1993)}]{36king1993theory}%
  \BibitemOpen
  \bibfield  {author} {\bibinfo {author} {\bibfnamefont {R.~D.}\ \bibnamefont {King-Smith}}\ and\ \bibinfo {author} {\bibfnamefont {D.}~\bibnamefont {Vanderbilt}},\ }\href {https://doi.org/10.1103/physrevb.47.1651} {\bibfield  {journal} {\bibinfo  {journal} {Phys. Rev. B}\ }\textbf {\bibinfo {volume} {47}},\ \bibinfo {pages} {1651} (\bibinfo {year} {1993})}\BibitemShut {NoStop}%
\bibitem [{\citenamefont {Taylor}\ \emph {et~al.}(2001{\natexlab{a}})\citenamefont {Taylor}, \citenamefont {Guo},\ and\ \citenamefont {Wang}}]{37taylor2001ab}%
  \BibitemOpen
  \bibfield  {author} {\bibinfo {author} {\bibfnamefont {J.}~\bibnamefont {Taylor}}, \bibinfo {author} {\bibfnamefont {H.}~\bibnamefont {Guo}},\ and\ \bibinfo {author} {\bibfnamefont {J.}~\bibnamefont {Wang}},\ }\href {https://doi.org/10.1103/PhysRevB.63.245407} {\bibfield  {journal} {\bibinfo  {journal} {Phys. Rev. B}\ }\textbf {\bibinfo {volume} {63}},\ \bibinfo {pages} {245407} (\bibinfo {year} {2001}{\natexlab{a}})}\BibitemShut {NoStop}%
\bibitem [{\citenamefont {Taylor}\ \emph {et~al.}(2001{\natexlab{b}})\citenamefont {Taylor}, \citenamefont {Guo},\ and\ \citenamefont {Wang}}]{38taylor2001ab}%
  \BibitemOpen
  \bibfield  {author} {\bibinfo {author} {\bibfnamefont {J.}~\bibnamefont {Taylor}}, \bibinfo {author} {\bibfnamefont {H.}~\bibnamefont {Guo}},\ and\ \bibinfo {author} {\bibfnamefont {J.}~\bibnamefont {Wang}},\ }\href {https://doi.org/10.1103/PhysRevB.63.121104} {\bibfield  {journal} {\bibinfo  {journal} {Phys. Rev. B}\ }\textbf {\bibinfo {volume} {63}},\ \bibinfo {pages} {121104} (\bibinfo {year} {2001}{\natexlab{b}})}\BibitemShut {NoStop}%
\bibitem [{\citenamefont {Datta}(1995)}]{39datta1997electronic}%
  \BibitemOpen
  \bibfield  {author} {\bibinfo {author} {\bibfnamefont {S.}~\bibnamefont {Datta}},\ }\href {https://doi.org/10.1017/cbo9780511805776} {\emph {\bibinfo {title} {Electronic transport in mesoscopic systems}}}\ (\bibinfo  {publisher} {Cambridge university press},\ \bibinfo {year} {1995})\BibitemShut {NoStop}%
\bibitem [{\citenamefont {Meir}\ and\ \citenamefont {Wingreen}(1992)}]{40meir1992landauer}%
  \BibitemOpen
  \bibfield  {author} {\bibinfo {author} {\bibfnamefont {Y.}~\bibnamefont {Meir}}\ and\ \bibinfo {author} {\bibfnamefont {N.~S.}\ \bibnamefont {Wingreen}},\ }\href {https://doi.org/10.1103/physrevlett.68.2512} {\bibfield  {journal} {\bibinfo  {journal} {Phys. Rev. Lett.}\ }\textbf {\bibinfo {volume} {68}},\ \bibinfo {pages} {2512} (\bibinfo {year} {1992})}\BibitemShut {NoStop}%
\bibitem [{\citenamefont {Kang}\ \emph {et~al.}(2020)\citenamefont {Kang}, \citenamefont {Jiang}, \citenamefont {Hao}, \citenamefont {Zhou}, \citenamefont {Zheng}, \citenamefont {Zhang},\ and\ \citenamefont {Zeng}}]{42kang2020giant}%
  \BibitemOpen
  \bibfield  {author} {\bibinfo {author} {\bibfnamefont {L.}~\bibnamefont {Kang}}, \bibinfo {author} {\bibfnamefont {P.}~\bibnamefont {Jiang}}, \bibinfo {author} {\bibfnamefont {H.}~\bibnamefont {Hao}}, \bibinfo {author} {\bibfnamefont {Y.}~\bibnamefont {Zhou}}, \bibinfo {author} {\bibfnamefont {X.}~\bibnamefont {Zheng}}, \bibinfo {author} {\bibfnamefont {L.}~\bibnamefont {Zhang}},\ and\ \bibinfo {author} {\bibfnamefont {Z.}~\bibnamefont {Zeng}},\ }\href {https://doi.org/10.1103/physrevb.101.014105} {\bibfield  {journal} {\bibinfo  {journal} {Phys. Rev. B}\ }\textbf {\bibinfo {volume} {101}},\ \bibinfo {pages} {014105} (\bibinfo {year} {2020})}\BibitemShut {NoStop}%
\bibitem [{\citenamefont {Tao}\ and\ \citenamefont {Wang}(2016)}]{43tao2016ferroelectricity}%
  \BibitemOpen
  \bibfield  {author} {\bibinfo {author} {\bibfnamefont {L.}~\bibnamefont {Tao}}\ and\ \bibinfo {author} {\bibfnamefont {J.}~\bibnamefont {Wang}},\ }\href {https://doi.org/10.1063/1.4941805} {\bibfield  {journal} {\bibinfo  {journal} {Appl. Phys. Lett.}\ }\textbf {\bibinfo {volume} {108}},\ \bibinfo {pages} {062903} (\bibinfo {year} {2016})}\BibitemShut {NoStop}%
\bibitem [{\citenamefont {Li}\ \emph {et~al.}(2019{\natexlab{a}})\citenamefont {Li}, \citenamefont {Li}, \citenamefont {Du}, \citenamefont {Wang}, \citenamefont {Gu}, \citenamefont {Zhang}, \citenamefont {He}, \citenamefont {Duan},\ and\ \citenamefont {Xu}}]{li2019intrinsic}%
  \BibitemOpen
  \bibfield  {author} {\bibinfo {author} {\bibfnamefont {J.}~\bibnamefont {Li}}, \bibinfo {author} {\bibfnamefont {Y.}~\bibnamefont {Li}}, \bibinfo {author} {\bibfnamefont {S.}~\bibnamefont {Du}}, \bibinfo {author} {\bibfnamefont {Z.}~\bibnamefont {Wang}}, \bibinfo {author} {\bibfnamefont {B.-L.}\ \bibnamefont {Gu}}, \bibinfo {author} {\bibfnamefont {S.-C.}\ \bibnamefont {Zhang}}, \bibinfo {author} {\bibfnamefont {K.}~\bibnamefont {He}}, \bibinfo {author} {\bibfnamefont {W.}~\bibnamefont {Duan}},\ and\ \bibinfo {author} {\bibfnamefont {Y.}~\bibnamefont {Xu}},\ }\href {https://doi.org/10.1126/sciadv.aaw5685} {\bibfield  {journal} {\bibinfo  {journal} {Sci. Adv.}\ }\textbf {\bibinfo {volume} {5}},\ \bibinfo {pages} {eaaw5685} (\bibinfo {year} {2019}{\natexlab{a}})}\BibitemShut {NoStop}%
\bibitem [{\citenamefont {Gong}\ \emph {et~al.}(2019)\citenamefont {Gong}, \citenamefont {Guo}, \citenamefont {Li}, \citenamefont {Zhu}, \citenamefont {Liao}, \citenamefont {Liu}, \citenamefont {Zhang}, \citenamefont {Gu}, \citenamefont {Tang}, \citenamefont {Feng} \emph {et~al.}}]{gong2019experimental}%
  \BibitemOpen
  \bibfield  {author} {\bibinfo {author} {\bibfnamefont {Y.}~\bibnamefont {Gong}}, \bibinfo {author} {\bibfnamefont {J.}~\bibnamefont {Guo}}, \bibinfo {author} {\bibfnamefont {J.}~\bibnamefont {Li}}, \bibinfo {author} {\bibfnamefont {K.}~\bibnamefont {Zhu}}, \bibinfo {author} {\bibfnamefont {M.}~\bibnamefont {Liao}}, \bibinfo {author} {\bibfnamefont {X.}~\bibnamefont {Liu}}, \bibinfo {author} {\bibfnamefont {Q.}~\bibnamefont {Zhang}}, \bibinfo {author} {\bibfnamefont {L.}~\bibnamefont {Gu}}, \bibinfo {author} {\bibfnamefont {L.}~\bibnamefont {Tang}}, \bibinfo {author} {\bibfnamefont {X.}~\bibnamefont {Feng}}, \emph {et~al.},\ }\href {https://doi.org/10.1088/0256-307X/36/7/076801} {\bibfield  {journal} {\bibinfo  {journal} {Chin. Phys. Lett.}\ }\textbf {\bibinfo {volume} {36}},\ \bibinfo {pages} {076801} (\bibinfo {year} {2019})}\BibitemShut {NoStop}%
\bibitem [{\citenamefont {Zhang}\ \emph {et~al.}(2019)\citenamefont {Zhang}, \citenamefont {Wang}, \citenamefont {Wang}, \citenamefont {Wei}, \citenamefont {Chen}, \citenamefont {Wang}, \citenamefont {Shi}, \citenamefont {Wang}, \citenamefont {Jia}, \citenamefont {Ouyang} \emph {et~al.}}]{zhang2019experimental}%
  \BibitemOpen
  \bibfield  {author} {\bibinfo {author} {\bibfnamefont {S.}~\bibnamefont {Zhang}}, \bibinfo {author} {\bibfnamefont {R.}~\bibnamefont {Wang}}, \bibinfo {author} {\bibfnamefont {X.}~\bibnamefont {Wang}}, \bibinfo {author} {\bibfnamefont {B.}~\bibnamefont {Wei}}, \bibinfo {author} {\bibfnamefont {B.}~\bibnamefont {Chen}}, \bibinfo {author} {\bibfnamefont {H.}~\bibnamefont {Wang}}, \bibinfo {author} {\bibfnamefont {G.}~\bibnamefont {Shi}}, \bibinfo {author} {\bibfnamefont {F.}~\bibnamefont {Wang}}, \bibinfo {author} {\bibfnamefont {B.}~\bibnamefont {Jia}}, \bibinfo {author} {\bibfnamefont {Y.}~\bibnamefont {Ouyang}}, \emph {et~al.},\ }\href {https://doi.org/https://doi.org/10.1021/acs.nanolett.9b04555} {\bibfield  {journal} {\bibinfo  {journal} {Nano Lett.}\ }\textbf {\bibinfo {volume} {20}},\ \bibinfo {pages} {709} (\bibinfo {year} {2019})}\BibitemShut {NoStop}%
\bibitem [{\citenamefont {Chen}\ \emph {et~al.}(2024)\citenamefont {Chen}, \citenamefont {Liu}, \citenamefont {Li}, \citenamefont {Tay}, \citenamefont {Taniguchi}, \citenamefont {Watanabe}, \citenamefont {Chan}, \citenamefont {Yan}, \citenamefont {Song}, \citenamefont {Cheng} \emph {et~al.}}]{chen2024even}%
  \BibitemOpen
  \bibfield  {author} {\bibinfo {author} {\bibfnamefont {B.}~\bibnamefont {Chen}}, \bibinfo {author} {\bibfnamefont {X.}~\bibnamefont {Liu}}, \bibinfo {author} {\bibfnamefont {Y.}~\bibnamefont {Li}}, \bibinfo {author} {\bibfnamefont {H.}~\bibnamefont {Tay}}, \bibinfo {author} {\bibfnamefont {T.}~\bibnamefont {Taniguchi}}, \bibinfo {author} {\bibfnamefont {K.}~\bibnamefont {Watanabe}}, \bibinfo {author} {\bibfnamefont {M.~H.}\ \bibnamefont {Chan}}, \bibinfo {author} {\bibfnamefont {J.}~\bibnamefont {Yan}}, \bibinfo {author} {\bibfnamefont {F.}~\bibnamefont {Song}}, \bibinfo {author} {\bibfnamefont {R.}~\bibnamefont {Cheng}}, \emph {et~al.},\ }\href {https://doi.org/https://doi.org/10.1021/acs.nanolett.4c01597} {\bibfield  {journal} {\bibinfo  {journal} {Nano Lett.}\ }\textbf {\bibinfo {volume} {24}},\ \bibinfo {pages} {8320} (\bibinfo {year} {2024})}\BibitemShut {NoStop}%
\bibitem [{\citenamefont {Li}\ and\ \citenamefont {Cheng}(2022)}]{li2022identifying}%
  \BibitemOpen
  \bibfield  {author} {\bibinfo {author} {\bibfnamefont {Y.-H.}\ \bibnamefont {Li}}\ and\ \bibinfo {author} {\bibfnamefont {R.}~\bibnamefont {Cheng}},\ }\href {https://doi.org/10.1103/PhysRevResearch.4.L022067} {\bibfield  {journal} {\bibinfo  {journal} {Phys. Rev. Research}\ }\textbf {\bibinfo {volume} {4}},\ \bibinfo {pages} {L022067} (\bibinfo {year} {2022})}\BibitemShut {NoStop}%
\bibitem [{\citenamefont {Zhan}\ \emph {et~al.}(2022)\citenamefont {Zhan}, \citenamefont {Yang}, \citenamefont {Luo}, \citenamefont {Zhang}, \citenamefont {Lou}, \citenamefont {Liu}, \citenamefont {Wu},\ and\ \citenamefont {Chang}}]{zhan2022spin}%
  \BibitemOpen
  \bibfield  {author} {\bibinfo {author} {\bibfnamefont {G.}~\bibnamefont {Zhan}}, \bibinfo {author} {\bibfnamefont {Z.}~\bibnamefont {Yang}}, \bibinfo {author} {\bibfnamefont {K.}~\bibnamefont {Luo}}, \bibinfo {author} {\bibfnamefont {D.}~\bibnamefont {Zhang}}, \bibinfo {author} {\bibfnamefont {W.}~\bibnamefont {Lou}}, \bibinfo {author} {\bibfnamefont {J.}~\bibnamefont {Liu}}, \bibinfo {author} {\bibfnamefont {Z.}~\bibnamefont {Wu}},\ and\ \bibinfo {author} {\bibfnamefont {K.}~\bibnamefont {Chang}},\ }\href {https://doi.org/10.1557/s43577-022-00381-8} {\bibfield  {journal} {\bibinfo  {journal} {MRS Bull.}\ }\textbf {\bibinfo {volume} {47}},\ \bibinfo {pages} {1177} (\bibinfo {year} {2022})}\BibitemShut {NoStop}%
\bibitem [{\citenamefont {Ding}\ \emph {et~al.}(2017)\citenamefont {Ding}, \citenamefont {Zhu}, \citenamefont {Wang}, \citenamefont {Gao}, \citenamefont {Xiao}, \citenamefont {Gu}, \citenamefont {Zhang},\ and\ \citenamefont {Zhu}}]{46ding2017prediction}%
  \BibitemOpen
  \bibfield  {author} {\bibinfo {author} {\bibfnamefont {W.}~\bibnamefont {Ding}}, \bibinfo {author} {\bibfnamefont {J.}~\bibnamefont {Zhu}}, \bibinfo {author} {\bibfnamefont {Z.}~\bibnamefont {Wang}}, \bibinfo {author} {\bibfnamefont {Y.}~\bibnamefont {Gao}}, \bibinfo {author} {\bibfnamefont {D.}~\bibnamefont {Xiao}}, \bibinfo {author} {\bibfnamefont {Y.}~\bibnamefont {Gu}}, \bibinfo {author} {\bibfnamefont {Z.}~\bibnamefont {Zhang}},\ and\ \bibinfo {author} {\bibfnamefont {W.}~\bibnamefont {Zhu}},\ }\href {https://doi.org/10.1038/ncomms14956} {\bibfield  {journal} {\bibinfo  {journal} {Nat. Commun.}\ }\textbf {\bibinfo {volume} {8}},\ \bibinfo {pages} {14956} (\bibinfo {year} {2017})}\BibitemShut {NoStop}%
\bibitem [{\citenamefont {Otrokov}\ \emph {et~al.}(2019)\citenamefont {Otrokov}, \citenamefont {Rusinov}, \citenamefont {Blanco-Rey}, \citenamefont {Hoffmann}, \citenamefont {Vyazovskaya}, \citenamefont {Eremeev}, \citenamefont {Ernst}, \citenamefont {Echenique}, \citenamefont {Arnau},\ and\ \citenamefont {Chulkov}}]{mbt-otrokov2019unique}%
  \BibitemOpen
  \bibfield  {author} {\bibinfo {author} {\bibfnamefont {M.}~\bibnamefont {Otrokov}}, \bibinfo {author} {\bibfnamefont {I.~P.}\ \bibnamefont {Rusinov}}, \bibinfo {author} {\bibfnamefont {M.}~\bibnamefont {Blanco-Rey}}, \bibinfo {author} {\bibfnamefont {M.}~\bibnamefont {Hoffmann}}, \bibinfo {author} {\bibfnamefont {A.~Y.}\ \bibnamefont {Vyazovskaya}}, \bibinfo {author} {\bibfnamefont {S.}~\bibnamefont {Eremeev}}, \bibinfo {author} {\bibfnamefont {A.}~\bibnamefont {Ernst}}, \bibinfo {author} {\bibfnamefont {P.~M.}\ \bibnamefont {Echenique}}, \bibinfo {author} {\bibfnamefont {A.}~\bibnamefont {Arnau}},\ and\ \bibinfo {author} {\bibfnamefont {E.~V.}\ \bibnamefont {Chulkov}},\ }\href {https://doi.org/10.1103/physrevlett.122.107202} {\bibfield  {journal} {\bibinfo  {journal} {Phys. Rev. Lett.}\ }\textbf {\bibinfo {volume} {122}},\ \bibinfo {pages} {107202} (\bibinfo {year} {2019})}\BibitemShut {NoStop}%
\bibitem [{\citenamefont {Yuan}\ \emph {et~al.}(2024)\citenamefont {Yuan}, \citenamefont {Dai}, \citenamefont {Liu},\ and\ \citenamefont {Zhao}}]{gr-yuan2024tunneling}%
  \BibitemOpen
  \bibfield  {author} {\bibinfo {author} {\bibfnamefont {J.}~\bibnamefont {Yuan}}, \bibinfo {author} {\bibfnamefont {J.-Q.}\ \bibnamefont {Dai}}, \bibinfo {author} {\bibfnamefont {Y.-Z.}\ \bibnamefont {Liu}},\ and\ \bibinfo {author} {\bibfnamefont {M.-W.}\ \bibnamefont {Zhao}},\ }\href {https://doi.org/10.1016/j.surfin.2024.103977} {\bibfield  {journal} {\bibinfo  {journal} {Surf. Interfaces}\ }\textbf {\bibinfo {volume} {46}},\ \bibinfo {pages} {103977} (\bibinfo {year} {2024})}\BibitemShut {NoStop}%
\bibitem [{\citenamefont {Li}\ \emph {et~al.}(2019{\natexlab{b}})\citenamefont {Li}, \citenamefont {Lu}, \citenamefont {Zhang}, \citenamefont {You}, \citenamefont {Su},\ and\ \citenamefont {Tsymbal}}]{48li2019spin}%
  \BibitemOpen
  \bibfield  {author} {\bibinfo {author} {\bibfnamefont {X.}~\bibnamefont {Li}}, \bibinfo {author} {\bibfnamefont {J.-T.}\ \bibnamefont {Lu}}, \bibinfo {author} {\bibfnamefont {J.}~\bibnamefont {Zhang}}, \bibinfo {author} {\bibfnamefont {L.}~\bibnamefont {You}}, \bibinfo {author} {\bibfnamefont {Y.}~\bibnamefont {Su}},\ and\ \bibinfo {author} {\bibfnamefont {E.~Y.}\ \bibnamefont {Tsymbal}},\ }\href {https://doi.org/10.1021/acs.nanolett.9b01506} {\bibfield  {journal} {\bibinfo  {journal} {Nano Lett.}\ }\textbf {\bibinfo {volume} {19}},\ \bibinfo {pages} {5133} (\bibinfo {year} {2019}{\natexlab{b}})}\BibitemShut {NoStop}%
\end{thebibliography}%
\end{document}